\PassOptionsToPackage{unicode}{hyperref}
\PassOptionsToPackage{hyphens}{url}
\PassOptionsToPackage{dvipsnames,svgnames,x11names}{xcolor}
\documentclass[
  12pt]{article}
  
\usepackage{amsmath,amssymb}
\usepackage{amsfonts}
\usepackage{algorithm, algorithmic}
\usepackage{newtxtext}
\usepackage{bm}
\usepackage{iftex}
\usepackage{enumitem}
\usepackage[subscriptcorrection]{newtxmath}
\usepackage{rotating}
\usepackage{multirow}

\newtheorem{theorem}{Theorem}
\newtheorem{assumption}{Assumption}

\newtheorem{remark}{Remark}

\newtheorem{corollary}{Corollary}
\newtheorem{proposition}{Proposition}

\newcommand{\diag}{\operatorname{diag}}
\newcommand{\ndiag}{\operatorname{ndiag}}
\newcommand{\plim}{\operatorname{plim}}
\newcommand{\sgn}{\operatorname{sgn}}
\ifPDFTeX
  \usepackage[T1]{fontenc}
  \usepackage[utf8]{inputenc}
  \usepackage{textcomp} 
\else 
  \usepackage{unicode-math}
  \defaultfontfeatures{Scale=MatchLowercase}
  \defaultfontfeatures[\rmfamily]{Ligatures=TeX,Scale=1}
\fi

\ifPDFTeX\else  
\fi
\IfFileExists{upquote.sty}{\usepackage{upquote}}{}
\IfFileExists{microtype.sty}{
  \usepackage[]{microtype}
  \UseMicrotypeSet[protrusion]{basicmath} 
}{}
\makeatletter
\@ifundefined{KOMAClassName}{
  \IfFileExists{parskip.sty}{%
    \usepackage{parskip}
  }{
    \setlength{\parindent}{0pt}
    \setlength{\parskip}{6pt plus 2pt minus 1pt}}
}{
  \KOMAoptions{parskip=half}}
\makeatother
\usepackage{xcolor}
\makeatletter
\ifx\paragraph\undefined\else
  \let\oldparagraph\paragraph
  \renewcommand{\paragraph}{
    \@ifstar
      \xxxParagraphStar
      \xxxParagraphNoStar
  }
  \newcommand{\xxxParagraphStar}[1]{\oldparagraph*{#1}\mbox{}}
  \newcommand{\xxxParagraphNoStar}[1]{\oldparagraph{#1}\mbox{}}
\fi
\ifx\subparagraph\undefined\else
  \let\oldsubparagraph\subparagraph
  \renewcommand{\subparagraph}{
    \@ifstar
      \xxxSubParagraphStar
      \xxxSubParagraphNoStar
  }
  \newcommand{\xxxSubParagraphStar}[1]{\oldsubparagraph*{#1}\mbox{}}
  \newcommand{\xxxSubParagraphNoStar}[1]{\oldsubparagraph{#1}\mbox{}}
\fi
\makeatother

\usepackage{longtable,booktabs,array}
\usepackage{calc} 
\usepackage{etoolbox}
\makeatletter
\patchcmd\longtable{\par}{\if@noskipsec\mbox{}\fi\par}{}{}
\makeatother
\IfFileExists{footnotehyper.sty}{\usepackage{footnotehyper}}{\usepackage{footnote}}
\makesavenoteenv{longtable}
\usepackage{graphicx}
\makeatletter
\def\maxwidth{\ifdim\Gin@nat@width>\linewidth\linewidth\else\Gin@nat@width\fi}
\def\maxheight{\ifdim\Gin@nat@height>\textheight\textheight\else\Gin@nat@height\fi}
\makeatother
\setkeys{Gin}{width=\maxwidth,height=\maxheight,keepaspectratio}
\makeatletter
\def\fps@figure{htbp}
\makeatother

\makeatletter
\@ifpackageloaded{caption}{}{\usepackage{caption}}
\AtBeginDocument{%
\ifdefined\contentsname
  \renewcommand*\contentsname{Table of contents}
\else
  \newcommand\contentsname{Table of contents}
\fi
\ifdefined\listfigurename
  \renewcommand*\listfigurename{List of Figures}
\else
  \newcommand\listfigurename{List of Figures}
\fi
\ifdefined\listtablename
  \renewcommand*\listtablename{List of Tables}
\else
  \newcommand\listtablename{List of Tables}
\fi
\ifdefined\figurename
  \renewcommand*\figurename{Figure}
\else
  \newcommand\figurename{Figure}
\fi
\ifdefined\tablename
  \renewcommand*\tablename{Table}
\else
  \newcommand\tablename{Table}
\fi
}
\@ifpackageloaded{float}{}{\usepackage{float}}
\floatstyle{ruled}
\@ifundefined{c@chapter}{\newfloat{codelisting}{h}{lop}}{\newfloat{codelisting}{h}{lop}[chapter]}
\floatname{codelisting}{Listing}

\makeatother
\makeatletter
\@ifpackageloaded{caption}{}{\usepackage{caption}}
\@ifpackageloaded{subcaption}{}{\usepackage{subcaption}}
\makeatother

\ifLuaTeX
  \usepackage{selnolig}  
\fi
\usepackage[]{natbib}
\usepackage{bookmark}

\IfFileExists{xurl.sty}{\usepackage{xurl}}{} 
\hypersetup{
  pdftitle={Title},
  pdfauthor={Author 1; Author 2},
  pdfkeywords={3 to 6 keywords, that do not appear in the title},
  colorlinks=true,
  linkcolor={blue},
  filecolor={Maroon},
  citecolor={Blue},
  urlcolor={Blue},
  pdfcreator={LaTeX via pandoc}}

\newcommand{\anon}{1}

\begin{document}

\def\spacingset#1{\renewcommand{\baselinestretch}%
{#1}\small\normalsize} \spacingset{1}


\if1\anon
{
  \title{\bf One-step group factor analysis via penalized least squares}
  \author{Xinbing Kong\\
    School of Statistics and Data Science, Nanjing Audit University,\\
    Nanjing 211815, China\\
    xinbingkong@126.com\\
    Xiaoying Pan\footnote{Corresponding author}\\
    School of Statistics and Data Science, Nanjing Audit University,\\
    Nanjing 211815, China\\
    panxiaoying0607@163.com\\
    Long Yu\\
    School of Statistics and Data Science,\\
    Shanghai University of Finance and Economics,\\
    Shanghai, 200433, China\\
    yulong@mail.shufe.edu.cn\\
    and\\
    Tong Zhang\\
    School of Statistics and Data Science, Nanjing Audit University,\\
    Nanjing, 211815, China\\
    zhangtongsta@163.com}
 \maketitle
} \fi

\if0\anon
{
  \bigskip
  \bigskip
  \bigskip
  \begin{center}
    {\LARGE\bf One-step group factor analysis via penalized least squares}
\end{center}
  \medskip
} \fi

\bigskip
\begin{abstract}
In this article, we revisit the problem of group factor analysis and propose a one-step penalized least squares method to estimate the factor loadings and factors in large-dimensional group factor models, offering a distinct alternative to the conventional two-step principal component approach. Our procedure originates from the equivalence between the group factor structure and the carefully tailored identification conditions. Leveraging this insight, we develop a tricky Lagrange multiplier formulation for a penalized least square loss function. This one-step optimization framework, combined with the fine-tuned penalty parameters, facilitates the direct derivation of central limit theorems for factor loadings, factor scores, common and local components, as well as the convergence rates for them. Our theory demonstrates that our one-step approach achieves the same rate as the two-step aggregated principal-component method and even the same limiting standard error in the balanced group panel case, but smaller limiting standard error than the two-step canonical correlation procedure. Extensive simulation studies justify the theory. Applications to U.S. house prices and CSI300 weekly returns confirm that our method identifies global factors and heterogeneous local patterns.
\end{abstract}

\noindent%
\vfill

\newpage
\spacingset{1.8} 

\section{Introduction}

The prevalence of high-dimensional data poses a fundamental challenge in modern statistics in effectively extracting its intrinsic structure while maintaining interpretability. Factor models have long served as essential tools toward the intrinsic structure learning. By projecting observed variables onto a smaller set of latent common factors, they facilitate dimensionality reduction and yield a structured representation of data. These models have consequently found widespread application across diverse fields such as finance, genomics, psychology, and the social sciences.
The classical factor models in \cite{2002Bai,2003Bai,2010Onatski,2012Onatski,2024He}, though powerful for dimension reduction and latent structure discovery, typically treat all variables as members of a homogeneous set, thus ignoring valuable prior grouping information. For example, the case study proposed by \cite{2019Wang} indicates that companies from the same industry are more likely to be grouped into the same category. This limitation has spurred the development of group factor models, which extend the traditional framework by explicitly incorporating group-specific latent factors alongside potential global common factors.

\ \ \ \ Group factor models have broad applicability across multiple disciplines. We illustrate this with three canonical examples. In international trade, let $y_{m,i,t}$ be the trade volume from country $m$ to country $i$ in year $t$. As discussed in \cite{2020Dhyne} and \cite{2022Bernard}, one global factor is worldwide advances in transportation technology, which affects all trade volumes among countries. The local factors encompass a range of regional elements that exert influence on specific country pairs, including export price levels. In the economic growth study \citep{2007Koren}, let $y_{m,i,t}$ be the growth rate of value-added per worker for industry $i$ in country $m$ in year $t$. \cite{2025Jin} show that events such as the COVID-19 pandemic or a global financial crisis serve as global factors, and country-specific factors affect all domestic industries in the corresponding country. In macroeconomics \citep{2016Andrade}, let $y_{m,i,t}$ be the inflation rate for product $i$ in location $m$ at date $t$. The global liquidity shocks exemplify a global factor, whereas local factors include a technology innovation for a particular product.

\ \ \ \ There are three main parallel streams of works in estimating the group factor models. The first is principal component (PC) estimation. \cite{2010Wang} presented a sequential PC procedure that iteratively refines estimates of global and local factors. However, the convergence of this iterative algorithm crucially relies on the quality of the initial estimates.
The second consists of Bayesian techniques, c.f., \cite{2003Kose, 2016Zhao}, but they suffer from heavy computational burdens, as noted by \cite{2016Breitung}. 
Another prominent strand of literature employs canonical correlation analysis (CCA) to estimate group factor models \citep{2016Breitung, 2019Andreou, 2023Choi, 2022Lin}. However, several limitations persist within this line of work. For example, the approach introduced by \cite{2016Breitung} assumes that the number of global factors is known. \cite{2019Andreou} developed a testing procedure based on the asymptotic distribution of the canonical correlation between factor spaces within a two-block framework. However, this approach relies on strong assumptions and is limited to settings that involve only two groups. \cite{2023Choi} proposed two consistent selection criteria based on average pairwise canonical correlations across all groups. However, if certain groups share similar local factor structures, pairwise canonical correlation analysis may incorrectly classify these common local factors as global, resulting in inconsistent estimation of global factors. This limitation has been confirmed in the simulation studies conducted by \cite{2025Hu}.
A generalized canonical correlation method was proposed by \cite{2022Lin}. Although this approach ensures valid estimation even when certain groups share similar local factors, it requires performing singular value decomposition on high-dimensional matrices, leading to an increase in computational burden.
Off the above three main streams of works, two most recent works are worthy of being mentioned. \cite{2023Chen} proposes projecting out estimated factors within the group to retain global factors. However, repeated matrix operations may degrade estimation efficiency. \cite{2025Hu} estimated global factors through the eigen-decomposition of a pooled projection matrix, aiming to maximize the average correlation between latent global factors and group-specific factors. Both methods belong to the class of two-step estimation approaches, first performing factor analysis separately within groups and then disentangling global and local factors. 

\ \ \ \ In this paper, we revisit the problem of group factor analysis and propose a novel estimation framework for high-dimensional group factor models that operates in only one step. We view group factor modeling as a fitting problem that uses a parametric structure to reconstruct the data panel. The parametric structure is a latent and identifiable framework under some constraints, which easily separates the global and local factors and identifies the loadings for the global and local factors, simultaneously. Therefore, this perspective naturally motivates our one-step estimation approach, the penalized least squares or a kind of penalized quasi-maximum likelihood procedure. The merits of this new method are obvious. First, parameter estimation and disentanglement of global and local factors are achieved jointly in one step. Second, this method can be easily extended to the nonlinear group factor analysis, such as the quantile group factor model which models the quantiles of each variable in the data panel with group factors. It can also be extended to data panel with variables of mixed types, e.g., simultaneous presence of categorical data and continuous Gaussian variables. As a first attempt and to avoid missing the basic idea, we focus on the penalized least square method in the present paper and leave extensions for future work. Third, as demonstrated later, the group factor analysis reduces to a problem of M-estimation with reference to a Lagrangian function with identification restrictions on the global and local factor loading matrices. This facilitates a direct derivation of the central limit theorems of the common components associated with the global factors. With carefully designed Lagrangian multipliers, the Hessian matrix of the Lagrangian function can be made positive definite asymptotically, making the second-order properties easily obtained.  

\ \ \ \ The remainder of this paper is organized as follows. Section \ref{section2} introduces the grouped factor model and the identification conditions, along with the proposed penalized estimation method and optimization algorithm. Section \ref{section3} presents the main theoretical results. Section \ref{section4} contains comprehensive simulation studies, demonstrating the advantages of our method over existing approaches. Section \ref{section5} features two empirical studies, concerning U.S. housing price growth and the weekly returns of the CSI 300 Index, respectively. Finally, Section \ref{section6} provides a discussion. Proofs of the theoretical results and other technical details are provided in the Supplementary Material.

\section{Model, Identification and Estimation} \label{section2}
\subsection{Group Factor Model}
Let $y_{m,i,t}$ be the observed data for the $i$th cross-section unit in group $m$ at time $t$. Consider the following model \citep{2025Hu}:
\begin{equation}
	y_{m,i,t} = \mathbf{\gamma}_{m,i}' \mathbf{G}_t + \mathbf{\lambda}_{m,i}' \mathbf{F}_{m,t} + e_{m,i,t}, \quad m=1,\dots,M,\, i=1,\dots,N_m,\, t=1,\dots,T
\end{equation}
where the vectors ${\mathbf{G}}_{t}$ and ${\mathbf{F}}_{m,t}$ are the global and local factors, with dimensions $r_0\times1$ and $r_m\times1$, respectively; the vectors ${\mathbf{\gamma }}_{m,i}$ and ${\mathbf{{\lambda }}}_{m,i}$ are the global and local factor loadings, with dimensions $r_0\times1$ and $r_m\times1$, respectively; ${e}_{m,i,t}$ is the idiosyncratic error.

\ \ \ \ In vector form, we have:
\begin{equation}\label{model1}
	\left[ \begin{matrix}
		{{\mathbf{Y}}_{1,t}}  \\
		{{\mathbf{Y}}_{2,t}}  \\
		\vdots   \\
		{{\mathbf{Y}}_{M,t}}  \\
	\end{matrix} \right]=\left[ \begin{matrix}
		{{\mathbf{\Gamma }}_{1}} & {{\mathbf{\Lambda }}_{1}} & \mathbf{0} & \cdots  & \mathbf{0}  \\
		{{\mathbf{\Gamma }}_{2}} & \mathbf{0} & {{\mathbf{\Lambda }}_{2}} & \cdots  & \mathbf{0}  \\
		\vdots  & \vdots  & \vdots  & \ddots  & \vdots   \\
		{{\mathbf{\Gamma }}_{M}} & \mathbf{0} & \mathbf{0} & \cdots  & {{\mathbf{\Lambda }}_{M}}  \\
	\end{matrix} \right]\left[ \begin{matrix}
		{{\mathbf{G}}_{t}}  \\
		{{\mathbf{F}}_{1,t}}  \\
		\vdots   \\
		{{\mathbf{F}}_{M,t}}  \\
	\end{matrix} \right]+\left[ \begin{matrix}
		{{\mathbf{e}}_{1,t}}  \\
		{{\mathbf{e}}_{2,t}}  \\
		\vdots   \\
		{{\mathbf{e}}_{M,t}}  \\
	\end{matrix} \right],
\end{equation}
where ${\mathbf{Y}}_{m,t}=[y_{m,1,t},\dots,y_{m,N_m,t}]'$ collects observations for $N_m$ variables in group $m$,  ${\mathbf{\Gamma}}_{m}=[\gamma_{m,1},\dots,\gamma_{m,N_m}]'$ and ${\mathbf{\Lambda }}_{m}=[\lambda_{m,1},\dots,\lambda_{m,N_m}]'$ are the matrices of global and local factor loadings of dimensions $N_m\times r_0$ and $N_m\times r_m$, respectively, and ${\mathbf{e}}_{m,t}=[e_{m,1,t},\dots,e_{m,N_{m},t}]'$ is the vector of error terms, with $m=1,\dots,M, t=1,\dots,T$. $r_0=0$ represents the absence of global factors. Let $N=\sum_{m=1}^{M}N_m$ and $r=\sum_{m=1}^{M}r_m$. Seen from model (\ref{model1}), the group factor model is a special factor model with $(r_0+r)$ factors that restricts the off-diagonal blocks of the submatrix (last $r$ columns) of the factor loading matrix to be zero matrices.  

\ \ \ \ Stacking the model over groups yields:
\begin{equation}\label{eq_time_model}
	\mathbf{Y}_t=\mathbf{\Psi}\mathbf{K}_t + \mathbf{e}_t=:\mathbf{\Gamma}\mathbf{G}_t+\mathbf{\Lambda}\mathbf{F}_t+\mathbf{e}_t,
\end{equation}
where $\mathbf{Y}_t=[\mathbf{Y}_{1,t}',\dots,\mathbf{Y}_{M,t}']'$ is the $N$-dimensional observation vector, $\mathbf{\Psi}=\left[\mathbf{\Gamma},\mathbf{\Lambda}\right]$ is the $N\times \left(r_0+r\right)$ loading matrix, comprising $\mathbf{\Gamma}=[\mathbf{\Gamma}_1',\dots,\mathbf{\Gamma}_M']'$ of dimension $N\times r_0$ and 
$\mathbf{\Lambda}=\diag(\mathbf{\Lambda}_1,\dots,\mathbf{\Lambda}_M)$ of dimension $N\times r$.
The latent factor vector and the error term are respectively $\mathbf{K}_t=[\mathbf{G}_t', \mathbf{F}_{t}']'$ of dimension $\left(r_0+r\right)$ and $\mathbf{e}_t=[\mathbf{e}_{1,t}',\dots,\mathbf{e}_{M,t}']'$ of dimension $N$, where $r$-dimensional vector $\mathbf{F}_t=[\mathbf{F}_{1,t}',\dots,\mathbf{F}_{M,t}']'$ collects the local factors.
We assume that the local factors are orthogonal to the global factors, both have zero mean, and the covariance matrix of $\mathbf{K}_t$ is  
\begin{equation}
	{{\Sigma }_{\mathbf{K}_t}}=\text{Var}\left[ \begin{matrix}
		{{\mathbf{G}}_{t}}  \\
		{{\mathbf{F}}_{t}}   \\
	\end{matrix} \right]=\left[ \begin{matrix}
		\Sigma _G & 0   \\
		0 & \Sigma _F \\
	\end{matrix} \right]=\text{Var}\left[ \begin{matrix}
		{{\mathbf{G}}_{t}}  \\
		{{\mathbf{F}}_{1,t}}  \\
		\vdots   \\
		{{\mathbf{F}}_{M,t}}  \\
	\end{matrix} \right]=\left[ \begin{matrix}
		\Sigma _G & 0 & \cdots  & 0  \\
		0 & {{\Sigma }_{11}} & \cdots  & {{\Sigma }_{1M}}  \\
		\vdots  & \vdots  & \ddots  & \vdots   \\
		0 & {{\Sigma }_{M1}} & \cdots  & {{\Sigma }_{MM}}  \\
	\end{matrix} \right].
\end{equation}

\ \ \ \ Stacking the factors over groups and time, we have:
\begin{equation}\label{eq_model}
	\mathbf{Y}=\mathbf{\Psi}\mathbf{K}+\mathbf{e},
\end{equation}
where $\mathbf{Y}=[\mathbf{Y}_{1},\dots,\mathbf{Y}_{T}], \mathbf{K}=[\mathbf{K}_1,\dots,\mathbf{K}_T]=[\mathbf{G}',\mathbf{F}']'$ and $\mathbf{e}=[\mathbf{e}_{1},\dots,\mathbf{e}_{T}]$ are matrices of dimensions $N\times T$, $\left(r_0+r\right)\times T$, and $N\times T$. $\mathbf{G}=[\mathbf{G}_1,\dots,\mathbf{G}_T]$ and $\mathbf{F}=[\mathbf{F}_1,\dots,\mathbf{F}_T]$ are matrices of dimensions $r_0\times T$ and $r\times T$.

\subsection{Identification Restrictions}
Like all factor models, the group factor model is identified only up to rotation matrices. To separately identify the global factors, local factors, and their corresponding loadings, we need some normalization conditions.

\ \ \ \ First, to identify the global factors and their loadings, we set $\mathbf{\Gamma}'\mathbf{\Gamma}/N=\sum_{m=1}^{M}\mathbf{\Gamma}_m'\mathbf{\Gamma}_m/N=\mathbf{I}_{r_0}$ and let ${\Sigma}_{G}$ be diagonal.  Second, to identify the local factors and their loadings, we set $\mathbf{\Lambda}\mathbf{\Lambda}'=\diag\{\mathbf{\Lambda}_1\mathbf{\Lambda}_1', \cdots, \mathbf{\Lambda}_M\mathbf{\Lambda}_M'\}=\diag\{N_1\mathbf{I}_{r_1}, \cdots, N_M\mathbf{I}_{r_M}\}$, and let $\Sigma_F$ be a diagonal matrix that amounts to saying that ${\Sigma}_{mm}$ are diagonal matrices for all $m=1,\dots, M$. Third, we assume the global and local factor loadings are orthogonal to each other, i.e., $\mathbf{\Gamma}'\mathbf{\Lambda}=\mathbf{0}$. We admit that when $M>2$, our condition rules out the setting where some pairs of groups share common local factors. However, we believe our procedure could be extended to this general case as long as positive-definiteness of $\Sigma_F$ is imposed. But for ease of exposition of our one-step architecture with direct optimization, we assume the above formally given identifiable conditions hold.   

\subsection{Least Squares Method} \label{2}
For model \eqref{model1}, the identification conditions for the loading matrix $\mathbf{\Psi}$ and the factor matrix $\mathbf{K}$ are summarized as follows. 
\begin{equation}\label{identification}
	\left\{\begin{array}{lll}
		\mathbf{K}\mathbf{K}'\text{ is diagonal matrix,}\\
		\mathbf{\Gamma}'\mathbf{\Lambda}=\mathbf{0},\\
		\frac{1}{N}\mathbf{\Gamma}'\mathbf{\Gamma}=\mathbf{I}_{r_0},\\
		\frac{1}{N_m}\mathbf{\Lambda}_m'\mathbf{\Lambda}_m=\mathbf{I}_{r_m},m=1,\dots,M.
	\end{array}
	\right.
\end{equation}

\ \ \ \ With the parametric model \eqref{model1}, it is natural to approximate the observed $y_{m,i,t}$'s by minimizing the square loss
\begin{equation*}
	L(\mathbf{\Psi},\mathbf{K})
	=\frac{1}{2}\|\mathbf{Y}-\mathbf{\Psi}\mathbf{K}\|_F^2
	=\frac{1}{2}\sum_{m=1}^{M}\sum_{i=1}^{N_m}\sum_{t=1}^{T}\left({y}_{m,i,t}-{\mathbf{{\gamma }'}}_{m,i}{\mathbf{G}}_{t}-{\mathbf{{\lambda }'}}_{m,i}{\mathbf{F}}_{m,t}\right)^2 \ \mbox{subject to} \ (\ref{identification}).
\end{equation*}

\ \ \ \ Our one-step estimate of the group factor model is the minimizer of the following penalized least square loss function (the Lagrangian function). 
\begin{equation}\label{optimization}
	\text{min}_{\left\{\mathbf{\Psi},\mathbf{K}\right\}}
	\mathcal{L}(\mathbf{\Psi},\mathbf{K})
	:= L(\mathbf{\Psi},\mathbf{K})
	+\mathbb{P}(\mathbf{\Psi},\mathbf{K}),
\end{equation}
where
\begin{align*}
	\mathbb{P}(\mathbf{\Psi},\mathbf{K}) =& bNT \left[ \frac{1}{8} \|\text{diag}(\frac{1}{N}\mathbf{\Gamma}'\mathbf{\Gamma}-\mathbf{I}_{r_0})\|_F^2
	+\frac{1}{2}\|\text{ndiag}(\frac{1}{N}\mathbf{\Gamma}'\mathbf{\Gamma})\|_F^2
	+\frac{1}{2}\|\text{ndiag}(\frac{1}{T}\mathbf{K}\mathbf{K}')\|_F^2
	\right.\\
	&\left.+\sum_{m=1}^{M}\left(\frac{1}{8}\|\text{diag}(\frac{1}{N_m}\mathbf{\Lambda}_m'\mathbf{\Lambda}_m-\mathbf{I}_{r_m})\|_F^2
	+\frac{1}{2}\|\text{ndiag}(\frac{1}{N_m}\mathbf{\Lambda}_{m}'\mathbf{\Lambda}_{m})\|_F^2\right)
	+\|\frac{1}{N}\mathbf{\Gamma}'\mathbf{\Lambda}\|_F^2\right],
\end{align*}
where $b$ is a positive Lagrange multiplier \citep{2026Kong, 2022Wang}, $\diag(\frac{1}{N}\mathbf{\Gamma}'\mathbf{\Gamma}-\mathbf{I}_{r_0})$ is a diagonal matrix consisting of the diagonal elements of $(\frac{1}{N}\mathbf{\Gamma}'\mathbf{\Gamma}-\mathbf{I}_{r_0})$, $\diag(\frac{1}{N_m}\mathbf{\Lambda}_m'\mathbf{\Lambda}_m-\mathbf{I}_{r_m})$ is defined in the same way. $\ndiag\left(\mathbf{K}\mathbf{K}'\right)$ is the upper triangular matrix consisting of nondiagonal elements of $\mathbf{K}\mathbf{K}'$, $\ndiag(\frac{1}{N}\mathbf{\Gamma}'\mathbf{\Gamma})$ and $\ndiag(\frac{1}{N_m}\mathbf{\Lambda}_{m}'\mathbf{\Lambda}_{m})$ are defined in the same way.

\ \ \ \ For the optimization problem \eqref{optimization}, let
$$
\theta=\left(\text{vec}\left(\mathbf{\Gamma}\right)',\text{vec}\left(\mathbf{\Lambda}_1\right)',\dots,\text{vec}\left(\mathbf{\Lambda}_M\right)',\text{vec}\left({\mathbf{G}}\right)',\text{vec}\left({\mathbf{F}}_{1}\right)',\dots,\text{vec}\left({\mathbf{F}}_{M}\right)'\right)'
$$
denote the parameter vector, and let 
$$
\theta^0=\left(\text{vec}\left(\mathbf{\Gamma}^0\right)',\text{vec}\left(\mathbf{\Lambda}_1^0\right)',\dots,\text{vec}\left(\mathbf{\Lambda}_M^0\right)',\text{vec}\left({\mathbf{G}^0}\right)',\text{vec}\left({\mathbf{F}}_{1}^0\right)',\dots,\text{vec}\left({\mathbf{F}}_{M}^0\right)'\right)'
$$ 
denote the corresponding true parameter vector, where $\mathbf{F}_m=[\mathbf{F}_{m,1},\dots,\mathbf{F}_{m,T}]$ and $\mathbf{F}_m^0=[\mathbf{F}_{m,1}^0,\dots,\mathbf{F}_{m,T}^0]$ are both of dimension $r_m\times T$.

\ \ \ \ The penalty term \(\mathbb{P}(\theta)\) is associated with condition \eqref{identification}. By strategically balancing the non-diagonal blocks of the scaled expectation of the Hessian matrix of \(L(\theta)\), the Hessian of the penalized loss function \(\mathcal{L}(\theta)\) is ensured to be locally positive-definite around the true parameter vector \(\theta^0\) asymptotically. To the best of our knowledge, this is the first time that the group factor analysis framework has been reformulated as a Lagrangian optimization problem, thereby facilitating the derivation of the central limit theorem for the estimated parameters. Although the present paper is concerned with the squared loss tailored for linear group factor analysis, it can be easily extended to the generalized linear or nonlinear group factor analysis by simply replacing the squared error loss (Gaussian likelihood function) with other (exponential family) likelihood functions or other loss functions (e.g. check loss for quantile factor analysis). This flexibility is not shared by the existing two-step approaches which depend on PCA techniques.   

\ \ \ \ To end this section, we introduce the following notation. Throughout the paper, the abbreviation ``w.p.a.1'' stands for ``with probability approaching 1''. For any vector, \(\|\cdot\|\) denotes its Euclidean norm. For any matrix \(\mathbf{A}\), let \(\rho_{\min}(\mathbf{\mathbf{A}})\) be its smallest eigenvalue; the symbols \(\|\mathbf{A}\|\), \(\|\mathbf{A}\|_F\), \(\|\mathbf{A}\|_1\), \(\|\mathbf{A}\|_\infty\), and \(\|\mathbf{A}\|_{\max}\) represent the spectral norm, Frobenius norm, \(L_1\)-norm, infinity norm, and max norm, respectively. Let sgn$(a)=1$ if $a\geq0$ and sgn$(a)=-1$ if $a<0$. Accordingly, $\text{sgn}(\mathbf{A})$ represents a diagonal matrix where the $i$-th diagonal element corresponds to $\text{sgn}(\mathbf{A}_{ii})$. For a vector $(a_1, \dots, a_p)$, let $\text{diag}(a_1, \dots, a_p)$ denote the $p \times p$ diagonal matrix with diagonal entries $a_1, \dots, a_p$. Conversely, when applied to a matrix $\mathbf{A}$, $\text{diag}(\mathbf{A})$ refers to the vector composed of the diagonal elements of $\mathbf{A}$. The $(i, j)$-th entry of a matrix $\mathbf{A}$ is written as $\mathbf{A}_{ij}$ (or $\mathbf{A}_{i,j}$). We denote the transpose of $\mathbf{A}$ by $\mathbf{A}^\top$ (or $\mathbf{A}'$), the trace by $\text{Tr}(\mathbf{A})$, and the rank by $\text{rank}(\mathbf{A})$. 
Let $\mathcal{S}(\theta) = \partial_\theta\mathcal{L}(\theta)$ be the score function. Let $\mathcal{H}(\theta) = \partial_{\theta^2} \mathcal{L}(\theta)$ be the Hessian matrix. The decomposition of $\mathcal{H}(\theta)$ and the expression of each component are presented in the Supplementary Material. 

\section{Theoretical Results}\label{section3}
In this section, we establish the convergence rates of the theoretical minimizers of \eqref{optimization} under sub-Gaussian tails of idiosyncratic errors, instead of the two-step iteration estimators in Algorithm 1 by \cite{2025Hu}. 

\ \ \ \ Let $B(\mathcal{D}) = \left\{ (\mathbf{\Gamma},\mathbf{\Lambda},\mathbf{G},\mathbf{F}) : \|\text{vec}(\mathbf{\Gamma})\|_\infty + \|\text{vec}(\mathbf{\Lambda})\|_\infty + \|\text{vec}(\mathbf{G})\|_\infty + \|\text{vec}(\mathbf{F})\|_\infty \leq \mathcal{D} \right\}$ for a constant $\mathcal{D }> 0$ large enough such that the true parameter $\theta^0$ lies in the interior of $B(\mathcal{D})$. Define $\hat{\mathbf{\Gamma}},\hat{\mathbf{\Lambda}},\hat{\mathbf{G}},\hat{\mathbf{F}}$
as the solution obtained by minimizing $\mathcal{L}(\theta)$ within $B(\mathcal{D})$. 
In this section, we present the asymptotic properties of the theoretical minimizers $\hat{\theta}$, defined as 
\begin{align*}
\hat{\theta}&=\left(\text{vec}\left({\hat{\mathbf{\Gamma}}}\right)',\text{vec}\left(\hat{\mathbf{\Lambda}}_1\right)',\dots,\text{vec}\left(\hat{\mathbf{\Lambda}}_M\right)',\text{vec}\left(\hat{\mathbf{G}}\right)',\text{vec}\left(\hat{\mathbf{F}}_1\right)',\dots,\text{vec}\left(\hat{\mathbf{F}}_M\right)'\right)'\\
	&=\text{arg }\text{min}_{\theta \in \Theta}\left\|\mathbf{Y}-\mathbf{\Gamma}\mathbf{G}-\mathbf{\Lambda}\mathbf{F}
    \right\|_F^2,
\end{align*}
where 
\begin{align*}
	\Theta=\{&\theta:\mathbf{\gamma}_{m,i}\in\mathcal{A}\subset\mathbb{R}^{r_0},\,\mathbf{\lambda}_{m,i}\in\mathcal{B}\subset\mathbb{R}^{r_m},\, \mathbf{K}_{t}\in\mathcal{K}\subset\mathbb{R}^{r_0+r}, \text{ for all }m\leq M,\,i\leq N_m,\,t\leq T,\,\\ 
    &\mathbf{\Gamma}'\mathbf{\Lambda}=\mathbf{0},\,
	\frac{1}{N}\mathbf{\Gamma}'\mathbf{\Gamma}=\mathbf{I}_{r_0},\, \frac{1}{N_m}\mathbf{\Lambda}'_m\mathbf{\Lambda}_m=\mathbf{I}_{r_m},\mathbf{K}\mathbf{K}'\text{ is a diagonal matrix}
	\}.
\end{align*}

\subsection{Consistency of the estimated factors and loadings}
To obtain the theoretical properties of $\hat{\theta}$, we assume that the following four assumptions hold. In the sequel, $C$ will be a generic constant that might vary from line to line.

\begin{assumption}\label{assume1}
	\begin{enumerate}[label=(\roman*)]
		\item $\mathcal{A}$, $\mathcal{B}$ and $\mathcal{K}$ are compact sets and $\theta^0\in\Theta$. 
		\item The factor matrix satisfies
		\begin{equation*}
			\frac{1}{T}\mathbf{K}^0\mathbf{K}^{0\prime}\rightarrow\Sigma,
			as\,T\rightarrow \infty,
		\end{equation*}
		where $\Sigma$ is a positive diagonal matrix with bounded distinct eigenvalues, with dimension $(r_0+r)\times (r_0+r)$. In particular, {$\frac{1}{T}\mathbf{G}\mathbf{G}'=\Sigma_{G}=\diag(\sigma_{Tg1},\dots,\sigma_{Tgr_0})$} with $\sigma_{Tg1}\geq\sigma_{Tg2}\cdots\geq\sigma_{Tgr_0}$, and $\sigma_{Tgi}\rightarrow \sigma_{gi}$ as $T\rightarrow \infty$ for $i=1,\dots,r_0$ with $\infty>\sigma_{g1}>\sigma_{g2}\dots>\sigma_{gr_0}>0$, $\frac{1}{T}\mathbf{F}_m\mathbf{F}_m'=\Sigma_{F_m}=\diag(\sigma_{Tf_m1},\dots,\sigma_{Tf_mr_m})$ for $m=1,\dots,M$, where $\sigma_{Tf_m1}\geq\sigma_{Tf_m2}\cdots\geq\sigma_{Tf_mr_m}$, and $\sigma_{Tf_mi}\rightarrow \sigma_{f_mi}$ as $T\rightarrow \infty$ for $i=1,\dots,r_m$ with $\infty>\sigma_{f_m1}>\sigma_{f_m2}\dots>\sigma_{f_mr_m}>0$.
	\end{enumerate}
\end{assumption}

\begin{assumption}\label{assume2}
	Given $\mathbf{K}^0$, $e_{m,i,t}$ are independent across $m,i$ and $t$.
\end{assumption}

\begin{assumption}\label{assume3}
	For $K>0$, $E(|e_{m,i,t}|^p|\mathbf{K}^0)\leq K^pp^{\frac{p}{2}}$ for all $p>1$, and $E(e_{m,i,t}|\mathbf{K}^0)=0$. 
\end{assumption}

\begin{assumption}\label{assume4}
	For all $m$, $\text{lim}_{N_m,N\rightarrow\infty}N/N_m=\alpha_m\leq C.$
\end{assumption}

\ \ \ \ Assumption \ref{assume1} is standard in large factor models. $\Sigma_G$ and $\Sigma_{F_m}$ are assumed to be diagonal matrices with distinctive diagonal elements so that the eigenvectors can be uniquely determined. This refers to \cite{2013Bai}, \cite{2023Chen&Fan} and \cite{2023He}. 
Assumption \ref{assume2} requires that the error terms are conditionally i.i.d. but can be dependent unconditionally. While this condition may appear restrictive, it is adopted for analytical convenience; see also \cite{2021ChenQuantile}.
Assumption 3 imposes sub-Gaussian tails on the idiosyncratic errors given the factor collection $\{\mathbf{K}^{0}\}$. This is made to streamline the theoretical derivations and is standard in high-dimensional statistical inference; see, for example, \cite{Wainwright_2019} and \cite{2024He}.
Assumption \ref{assume4} is necessary in the group factor model, as it imposes boundedness on the average number of $\{m\}$. Without this condition, the number of local factors $r=\sum_{m=1}^Mr_m$ may increase with the sample size, which contradicts the dimension-reduction philosophy underlying factor models and leads to the loss of a low-rank structure in the covariance matrix of the observed variables. Similar assumptions can be found in Assumption E1 of \cite{2025Hu} and Assumption E of \cite{2022Lin}.

\ \ \ \ The following theorem presents the asymptotic property of the estimators of {$\mathbf{\Gamma}, \mathbf{\Lambda}_m,\mathbf{G}$ and $\mathbf{F}_m$} for $m=1,\dots,M$ in terms of Frobenius norm.

\begin{theorem}\label{theorem1}
	{Let $\mathbf{S}_1=\sgn(\frac{1}{T}\hat{\mathbf{G}}\mathbf{G}^{0'})$, $\mathbf{S}_{2,m}=\sgn(\frac{1}{T}\hat{\mathbf{F}}_m\mathbf{F}_{m}^{0\prime})$. Then, under Assumptions \ref{assume1}--\ref{assume4}, 
	\begin{equation*}
		\frac{1}{N}\|\hat{\mathbf{\Gamma}}-\mathbf{\Gamma}^{0}\mathbf{S}_1\|^2_F
		+\frac{1}{N_m}\|\hat{\mathbf{\Lambda}}_m-\mathbf{\Lambda}_{m}^0\mathbf{S}_{2,m}\|^2_F
		+\frac{1}{T}\|\hat{\mathbf{G}}-\mathbf{S}_1\mathbf{G}^{0}\|^2_F
		+\frac{1}{T}\|\hat{\mathbf{F}}_m-\mathbf{S}_{2,m}\mathbf{F}_{m}^0\|^2_F
		=O_p(\delta^{-1}),
	\end{equation*}
	where $\delta=\min\{N,T\}$.} 
\end{theorem}

\ \ \ \ In Theorem \ref{theorem1}, the presence of the sign matrices \( \mathbf{S}_1 \) and \( \mathbf{S}_2 \) stems from the inherent sign ambiguity of factors and loadings, i.e., multiplying a factor and its corresponding loadings by \(-1\) leaves the factor structure unchanged. Theorem \ref{theorem1} establishes the consistency of the estimated parameters with the true parameters. This large-sample property provides a theoretical foundation for subsequent applications.

\ \ \ \ In comparison, the convergence rates of the factors and loadings estimated via the aggregated projection method \citep{2025Hu} and generalised canonical correlation estimation \citep{2022Lin} are $O_p(\min\{N_1,\dots,N_M,T\})^{-1}$. 
Under our Assumption \ref{assume4}, where $N \asymp N_m$, we obtain the same convergence rate as that in the existing literature.

\subsection{Asymptotic distributions of the estimated factors and loadings}
This section discusses the asymptotic distributions of the estimated factors and loadings. Under Assumptions \ref{assume1}--\ref{assume4}, the global and local factors can be consistently estimated. To derive the asymptotic distributions of the estimators, we need one more assumption.  

\begin{assumption}\label{assume5}
		For some positive definite matrices $\Omega_{\gamma, mi}$, $\Omega_{\lambda, mi}$, $\Omega_{G, mt}$ and $\Omega_{F, mt}$,
		\begin{equation*}
			\frac{1}{\sqrt{T}}\sum_{t=1}^{T}\mathbf{G}_t^0e_{m,i,t}\xrightarrow{d}N\left(\mathbf{0},\Omega_{\gamma, mi}\right),
			\quad
			\frac{1}{\sqrt{T}}\sum_{t=1}^{T}\mathbf{F}_{m,t}^0e_{m,i,t}\xrightarrow{d}N\left(\mathbf{0},\Omega_{\lambda, mi}\right),
		\end{equation*}
		where
		\begin{equation*}
			\Omega_{\gamma, mi}= \plim_{T \to \infty} \frac{1}{T} \sum_{s=1}^{T} \sum_{t=1}^{T} E\left( \mathbf{G}_s^0 \mathbf{G}_t^{0\prime} e_{m,i,s} e_{m,i,t} \right),
		\end{equation*}
        \begin{equation*}
            \Omega_{\lambda, mi} = \plim_{T \to \infty} \frac{1}{T} \sum_{s=1}^{T} \sum_{t=1}^{T} E\left( \mathbf{F}_{m,s}^0 \mathbf{F}_{m,t}^{0\prime} e_{m,i,s} e_{m,i,t} \right).
        \end{equation*}
		\begin{equation*}
			\frac{1}{\sqrt{N_m}}{\mathbf{\Gamma}_m^0}'\mathbf{e}_{m,t}\xrightarrow{d}N\left(\mathbf{0},\Omega_{G, mt}\right),
			\quad
			\frac{1}{\sqrt{N_m}}{\mathbf{\Lambda}_m^0}'\mathbf{e}_{m,t}\xrightarrow{d}N\left(\mathbf{0},\Omega_{F, mt}\right),
		\end{equation*}
		where
		\begin{equation*}
			\Omega_{G, mt}= \plim_{N_m \to \infty} \frac{1}{N_m} \sum_{i=1}^{N_m} \sum_{j=1}^{N_m} \gamma_{m,i}^0\gamma_{m,j}^{0\prime}E\left(e_{m,i,t} e_{m,j,t} \right), 
		\end{equation*}
		\begin{equation*}
			\Omega_{F,mt}= \plim_{N_m \to \infty} \frac{1}{N_m} \sum_{i=1}^{N_m} \sum_{j=1}^{N_m} \lambda_{m,i}^0\lambda_{m,j}^{0\prime}E\left(e_{m,i,t} e_{m,j,t} \right).
		\end{equation*}
\end{assumption}

\ \ \ \ Assumption \ref{assume5} is designed to derive the asymptotic distributions of the estimated factors and loadings. 
Without loss of generality, when the errors exhibit cross-sectional and serial dependence, additional assumptions are required to render them weakly correlated, see, for example, the representative works by \cite{2025Hu}, \cite{2023Chen} and \cite{2022Lin}.

\ \ \ \ To estimate the limiting distribution of the parameters, we need to compute the score function $\mathcal{S}\left(\theta\right)$ and the Hessian matrix $\mathcal{H}\left(\theta\right)$ of the objective function. By definition,
\begin{equation*}
    \mathcal{S}\left(\theta\right)=\partial_\theta\mathcal{L}\left(\theta\right)=\partial_\theta L\left(\theta\right)+\partial_\theta \mathbb{P}\left(\theta\right):=\mathcal{S}_L\left(\theta\right)+\mathcal{S}_P\left(\theta\right),
\end{equation*}
where
\begin{align*}
		\mathcal{S}_L(\theta)& =  \Big[  -\text{vec}\left(\left(\mathbf{Y}-\mathbf{\Gamma}\mathbf{G}-\mathbf{\Lambda}\mathbf{F}\right)\mathbf{G}'\right)',
		\dots, -\text{vec}\left(\left(\mathbf{Y}_m-\mathbf{\Gamma}_m\mathbf{G}-\mathbf{\Lambda}_m\mathbf{F}_m\right)\mathbf{F}_m'\right)',\dots, \\
		-\sum_{m=1}^{M}&\text{vec}\left(\mathbf{\Gamma}_m'\left(\mathbf{Y}_m-\mathbf{\Gamma}_m\mathbf{G}-\mathbf{\Lambda}_m\mathbf{F}_m\right)\right)',
		\dots, -\text{vec}\left(\mathbf{\Lambda}_m'\left(\mathbf{Y}_m-\mathbf{\Gamma}_m\mathbf{G}-\mathbf{\Lambda}_m\mathbf{F}_m\right)\right)',\dots
		\Big]'.
	\end{align*}

\ \ \ \ By Assumption \ref{assume1}, the true parameter $\theta^0$ satisfies the identification constraint \eqref{identification}. It follows that $\mathcal{S}_P\left(\theta^0\right)=0$. 
For $\mathcal{H}\left(\theta\right)$,
\begin{equation*}
    \mathcal{H}\left(\theta\right)=\partial_{\theta^2} \mathcal{L}(\theta) = \mathcal{H}_L\left(\theta\right) + \mathcal{J}_L\left(\theta\right) + \mathcal{H}_P\left(\theta\right),
\end{equation*}
where \( \mathcal{H}_L\left(\theta\right) \), \( \mathcal{J}_L\left(\theta\right) \) and \( \mathcal{H}_P\left(\theta\right) \) admit block structures with respect to the parameters. We relegate the explicit definitions of all block matrices to the Supplementary Material.
At the true value \( \theta^0 \), we denote in short $\mathcal{S}\left(\theta^0\right):=\mathcal{S}$, $\mathcal{S}_L\left(\theta^0\right):=\mathcal{S}_L$, $\mathcal{H}\left(\theta^0\right):=\mathcal{H}$, $\mathcal{H}_L\left(\theta^0\right):=\mathcal{H}_L$.

\ \ \ \ By proceeding to a higher-order expansion of the first-order condition, we obtain
\[
0 = \mathcal{S}(\hat{\theta}) = \mathcal{S} + \mathcal{H}(\hat{\theta} - \theta^0) + \frac{1}{2} E,
\]
where \(E\) is a residual matrix whose explicit form is provided in the Supplementary Material. Consequently,
\[
\hat{\theta} - \theta^0 = -\mathcal{H}^{-1}\mathcal{S} - \frac{1}{2}\mathcal{H}^{-1}E,
\]
\[
\hat{\gamma}_{mi} - \gamma_{mi}^0 = \left[\widehat{\theta} - \theta^0\right]_{\gamma mi} = -\left[\mathcal{H}^{-1}\mathcal{S}\right]_{ \gamma mi} - \frac{1}{2}\left[\mathcal{H}^{-1}E\right]_{\gamma mi},
\]
where $\left[\cdot\right]_{\gamma mi}$ denotes the $i$-th vector of the $m$-th block of $\left[\mathbf{I}_{Nr_0},\mathbf{0}_{Nr_0\times(\sum_{m=1}^MN_mr_m+(r_0+r)T)}\right]\left[\cdot\right]$, 
$\left[\cdot\right]_{\lambda mi}$ is the $i$-th vector of the $m$-th block of $\Big[\mathbf{0}_{(\sum_{m=1}^MN_mr_m)\times Nr_0},\mathbf{I}_{\sum_{m=1}^MN_mr_m}, \\
\mathbf{0}_{(\sum_{m=1}^MN_mr_m)\times (r_0+r)T}\Big]\left[\cdot\right]$, 
$\left[\cdot\right]_{g t}$ is the $t$-th vector of $\Big[\mathbf{0}_{r_0T\times (Nr_0+\sum_{m=1}^MN_mr_m)},
\mathbf{I}_{r_0T},\mathbf{0}_{r_0T\times rT}\Big]\left[\cdot\right]$, and 
$\left[\cdot\right]_{f mt}$ is the $t$-th vector of $m$-th block of $\left[\mathbf{0}_{rT\times (Nr_0+\sum_{m=1}^MN_mr_m+r_0T)},\mathbf{I}_{rT}\right]\left[\cdot\right]$.

\ \ \ \ By exploiting the local landscape property of \( \mathcal{H} \), we demonstrate in the Supplementary Material that
\begin{equation}\label{eqS}
    \left[\mathcal{H}^{-1} \mathcal{S}\right]_{\gamma mi} = \left[\mathcal{H}_{L11}^{-1} \mathcal{S}_\gamma\right]_{mi} + O_p\left((N T)^{-1/2}\right).
\end{equation}
where
$\mathcal{S}_\gamma=\left[\mathbf{I}_{Nr_0},\mathbf{0}_{Nr_0\times(\sum_{m=1}^MN_mr_m+(r_0+r)T)}\right]\mathcal{S}_L$, $\mathcal{S}_\lambda=\Big[\mathbf{0}_{(\sum_{m=1}^MN_mr_m)\times Nr_0},\mathbf{I}_{\sum_{m=1}^MN_mr_m},\\
\mathbf{0}_{(\sum_{m=1}^MN_mr_m)\times (r_0+r)T}\Big]\mathcal{S}_L$, $\mathcal{S}_g=\Big[\mathbf{0}_{r_0T\times (Nr_0+\sum_{m=1}^MN_mr_m)},
\mathbf{I}_{r_0T},\mathbf{0}_{r_0T\times rT}\Big]\mathcal{S}_L$, and $\mathcal{S}_f=\\ \Big[\mathbf{0}_{rT\times (Nr_0+\sum_{m=1}^MN_mr_m+r_0T)},
\mathbf{I}_{rT}\Big]\mathcal{S}_L$.

\ \ \ \ The rationale behind \eqref{eqS} stems from the nearly block-diagonal structure of \( \mathcal{H} \) induced by the Lagrangian penalty \( \mathbb{P}(\theta) \). More precisely, the entries lying on the diagonal blocks of \( \mathcal{H} \) dominate those in the off-diagonal blocks by a substantial margin. This particular pattern implies that when expanding \( [\mathcal{H}^{-1}S]_{\gamma mi} \), the contributions arising from the off-diagonal blocks are collectively of order $O_p \bigl( (N T)^{-1/2} \bigr)$,
 whereas \( \|[\mathcal{H}^{-1}E]_{\gamma mi}\| =  O_p\left( \frac{\left(\ln \left(NT\right)\right)^{3/2}}{\delta^2} \right) \).  Both of these terms are asymptotically negligible relative to the leading term on the right-hand side of \eqref{eqS}.

\begin{theorem}\label{theorem2}
	Under Assumptions \ref{assume1}--\ref{assume5}, 
	$$\sqrt{T}\left(\hat{\gamma}_{m,i} - \gamma_{m,i}^0 \right)\rightarrow N(\mathbf{0},\Sigma_G^{-1}\Omega_{\gamma, mi}\Sigma_G^{-1}), 
    $$
	$$\sqrt{T}\left(\hat{\lambda}_{m,i} - \lambda_{m,i}^0 \right)\rightarrow N(\mathbf{0},\Sigma_{F _m}^{-1}\Omega_{\lambda, mi}\Sigma_{F_m}^{-1}),
    $$
	$$\sqrt{N}\left(\hat{\mathbf{G}}_{t} - \mathbf{G}_{t}^0 \right)\rightarrow N(\mathbf{0},\sum_{m=1}^{M}\frac{1}{\alpha_m}\Omega_{G, mt}),
    $$
	$$\sqrt{N_m}\left(\hat{\mathbf{F}}_{m,t} - \mathbf{F}_{m,t}^0 \right)\rightarrow N(\mathbf{0},\Omega_{F, mt}),
   $$
\end{theorem}
where $\Sigma_G$ and $\Sigma_{F_m}$ are defined in Assumption \ref{assume1}, $\alpha_m$ is defined in Assumption \ref{assume4}, $\Omega_{\gamma, mi}$, $\Omega_{\lambda, mi}$, $\Omega_{G,mt}$ and $\Omega_{F,mt}$ are defined in Assumption \ref{assume5}.
\begin{remark}
	The asymptotic variances of $\hat{\gamma}_{m,i}, \hat{\lambda}_{m,i}, \hat{\mathbf{G}}_t, \hat{\mathbf{F}}_{m,t}$ can be consistently estimated, respectively, by
	$$
	\hat{\text{var}}_{\gamma}=T\left[\sum_{t=1}^{T}\hat{\mathbf{G}}_t\hat{\mathbf{G}}_t'\right]^{-1}\left[\sum_{t=1}^{T}\hat{e}_{m,i,t}^2\hat{\mathbf{G}}_t\hat{\mathbf{G}}_t'\right]\left[\sum_{t=1}^{T}\hat{\mathbf{G}}_t\hat{\mathbf{G}}_t'\right]^{-1},
	$$
	$$
	\hat{\text{var}}_{\lambda}=T\left[\sum_{t=1}^{T}\hat{\mathbf{F}}_{m,t}\hat{\mathbf{F}}_{m,t}'\right]^{-1}\left[\sum_{t=1}^{T}\hat{e}_{m,i,t}^2\hat{\mathbf{F}}_{m,t}\hat{\mathbf{F}}_{m,t}'\right]\left[\sum_{t=1}^{T}\hat{\mathbf{F}}_{m,t}\hat{\mathbf{F}}_{m,t}'\right]^{-1},
	$$
	$$
	\hat{\text{var}}_{g}=N\left[\sum_{m=1}^{M}\hat{\mathbf{\Gamma}}_m'\hat{\mathbf{\Gamma}}_m\right]^{-1}\left[\sum_{m=1}^{M}\frac{1}{\alpha_M^2}\sum_{i=1}^{N_m}\hat{e}_{m,i,t}^2\hat{\mathbf{\gamma}}_{m,i}\hat{\mathbf{\gamma}}_{m,i}'\right]\left[\sum_{m=1}^{M}\hat{\mathbf{\Gamma}}_m'\hat{\mathbf{\Gamma}}_m\right]^{-1},
	$$
    and 
	$$
	\hat{\text{var}}_{f}=N_m\left[\hat{\mathbf{\Lambda}}_m'\hat{\mathbf{\Lambda}}_m\right]^{-1}\left[\sum_{i=1}^{N_m}\hat{e}_{m,i,t}^2\hat{\mathbf{\lambda}}_{m,i}\hat{\mathbf{\lambda}}_{m,i}'\right]\left[\hat{\mathbf{\Lambda}}_m'\hat{\mathbf{\Lambda}}_m\right]^{-1}. 
	$$
\end{remark}

\ \ \ \ For the estimated common and local components, we soon have the following corollary. 
\begin{corollary}\label{corollary 1}
	Under Assumptions \ref{assume1}--\ref{assume5}, 
	\begin{equation*}
		\hat{\gamma}_{m,i}'\hat{\mathbf{G}}_{t}-\gamma_{m,i}^{0\prime}\mathbf{G}_{t}^0
		\rightarrow
		N\left(0, \frac{1}{N}\gamma_{m,i}^{0\prime}\sum_{m=1}^M\left(\frac{1}{\alpha_m}\Omega_{G, mt}\right)\gamma_{m,i}^{0}+
		\frac{1}{T}\mathbf{G}_t^{0\prime}\Sigma_{G}^{-1}\Omega_{\gamma, mi}\Sigma_{G}^{-1}\mathbf{G}_t^0
		\right),
	\end{equation*}
	\begin{equation*}
		\hat{\lambda}_{m,i}'\hat{\mathbf{F}}_{m,t}-\lambda_{m,i}^{0\prime}\mathbf{F}_{m,t}^0
		\rightarrow
		N\left(0,\frac{1}{T}\mathbf{F}_{m,t}^{0\prime}\Sigma_{F_m}^{-1}\Omega_{\lambda, mi}\Sigma_{F_m}^{-1}\mathbf{F}_{m,t}^0
		+\frac{1}{N_m}\lambda_{m,i}^{0\prime}\Omega_{F, mt}\lambda_{m,i}^{0}\right).
	\end{equation*}
\end{corollary}

\ \ \ \ To make comparison, we provide the central limit theorem for the estimated common and local components of \cite{2025Hu} (with upper tilde on the parameters to denote), that is not given in the reference paper.
\begin{proposition}\label{proposition 1}
    Under Assumptions A-F of \cite{2025Hu}, as $N_1, \dots, N_M, T\rightarrow \infty$, if $\frac{\sqrt{N}}{T} \rightarrow 0$ and $\frac{\sqrt{T}}{N} \rightarrow 0$. 
    \begin{equation*}
        \tilde{\gamma}_{m,i}'\tilde{\mathbf{G}}_t-\gamma_{m,i}^{0\prime}\mathbf{G}_t^0
        \rightarrow
        N\left(
        0,\frac{1}{M^2}\frac{1}{N}\gamma_{m,i}^{0\prime}\mathbb{C}\mathbb{V}_t^{(1)}\mathbb{C}'\gamma_{m,i}^0+\frac{1}{T}\mathbf{G}_t^{0\prime}\boldsymbol{\Sigma}_{G}^{-1}\mathbb{V}_{m,i}^{(2)}\boldsymbol{\Sigma}_{G}^{-1}\mathbf{G}_{t}^0
        \right),
    \end{equation*}
    and 
    \begin{align*}
		&\tilde{\lambda}_{m,i}'\tilde{\mathbf{F}}_{m,t}-\lambda_{m,i}^{0\prime}\mathbf{F}_{m,t}^0\rightarrow\\
		&
		N\left(0,\frac{1}{T}\mathbf{F}_{m,t}^{0\prime}\boldsymbol{\Sigma}_{F_m}^{-1}\mathbb{V}_{m,i}^{(3)}\boldsymbol{\Sigma}_{F_m}^{-1}\mathbf{F}_{m,t}^0+
		\frac{1}{N_m}\lambda_{m,i}^{0\prime}\mathbf{H}_m^{-1\prime}\boldsymbol{\Upsilon}_m^{-1} \mathbf{H}_m' \boldsymbol{\Sigma}_{mm} \mathbb{C}_1 \mathbb{V}_t^{(1)} \mathbb{C}_1' \boldsymbol{\Sigma}_{mm} \mathbf{H}_m \boldsymbol{\Upsilon}_m^{-1} \mathbf{H}_m^{-1}\lambda_{m,i}^0\right),
	\end{align*}
\end{proposition}
where $\mathbb{C}$ is defined in Theorem 5, $\mathbb{C}_1$ is defined in Theorem 6, $\mathbf{H}_m$ is defined in Theorem 4, $\mathbb{V}_t^{(1)}$ is defined in Assumption F3, $\mathbb{V}_{m,i}^{(2)}$ and $\mathbb{V}_{m,i}^{(3)}$ are defined in Assumption F4 of \cite{2025Hu}.

\ \ \ \ Theorem \ref{theorem2} and Corollary \ref{corollary 1} establish the central limit theorems for the quasi-likelihood method of the group factor model. Now, we compare the limiting variances of the common components with existing methods since the global factors are learned by transferring information from different groups. The limiting variance of the common component of \cite{2025Hu}'s estimate is $\frac{1}{M^2}\frac{1}{N}\gamma_{m,i}^{0\prime}\mathbb{C}\mathbb{V}_t^{(1)}\mathbb{C}'\gamma_{m,i}^0+\frac{1}{T}\mathbf{G}_t^{0\prime}\boldsymbol{\Sigma}_{G}^{-1}\mathbb{V}_{m,i}^{(2)}\boldsymbol{\Sigma}_{G}^{-1}\mathbf{G}_{t}^0$
, and ours is $\frac{1}{N}\gamma_{m,i}^{0\prime}\sum_{m=1}^M\left(\frac{1}{\alpha_m}\Omega_{G, mt}\right)\gamma_{m,i}^{0}+\frac{1}{T}\mathbf{G}_t^{0\prime}\Sigma_{G}^{-1}\Omega_{\gamma, mi}\Sigma_{G}^{-1}\mathbf{G}_t^0$. We proceed with a term-by-term comparison. 
For the first component, we assume that the idiosyncratic errors are mutually independent across groups, so that $\mathbb{V}_t^{(1)}$ in \cite{2025Hu} is block-diagonal with zero off-diagonal blocks. Consequently, the quadratic form $\mathbb{C}\mathbb{V}_t^{(1)}\mathbb{C}^{\prime}$ simplifies to 
\begin{equation*}
\mathbb{C}\mathbb{V}_t^{(1)}\mathbb{C}^{\prime}=\sum_{m=1}^M\alpha_m\mathbb{I}_m\Sigma_{{\Theta}_m}^{-1}\mathbb{V}_{mm,t}^{(1)}\Sigma_{{\Theta}_m}^{-1}\mathbb{I}_m',
\end{equation*}
where $\mathbb{V}_{mm,t}^{(1)}=\lim\frac{1}{N_m}\Theta_m'E[e_{m,t}e_{m,t}']\Theta_m$ and $\mathbb{I}_m=[\mathbf{I}_{r_0},\mathbf{0}_{r_0\times r_m}]$.
Under our identification assumption, this further reduces to $\mathbb{C}\mathbb{V}_t^1\mathbb{C}^{\prime}=\sum_{m=1}^M\alpha_m\Omega_{G,mt}$.
For a balanced panel with $N_m=N/M$ for all $m$, we have $\alpha_m=M$, and hence the first term of the variance expression is identical for both \cite{2025Hu}'s estimate and ours.
For the second term, $\mathbb{V}_{m,i}^{(2)}=\Omega_{\gamma, mi}$. Therefore, under our assumptions, \cite{2025Hu}'s APM method and our one-step least square method yield the same asymptotic variance for the estimated common components.

\ \ \ \ We also compare with \cite{2022Lin}'s generalised canonical correlation method. From Corollary 1 of \cite{2022Lin}, the asymptotic variance of the common component is \(\frac{1}{M^2}\frac{1}{N}\gamma_{i,j}^{0\prime}\mathcal{I}'\mathbb{C}^0\mathbb{D}_t^{(1)}\mathbb{C}^{0\prime}\mathcal{I}\gamma_{i,j}^0+\frac{1}{T}\mathbf{G}_t^{0\prime}\Sigma_{G}^{-1}\mathbb{D}_{i,j}^{(2)}\Sigma_G^{-1}\mathbf{G}_{t}^0\). The first term satisfies $\mathcal{I}'\mathbb{C}^0\mathbb{D}_t^{(1)}\mathbb{C}^{0\prime}\mathcal{I}=\sum_{m=1}^M\alpha_m\Omega_{G, mt}$. 
Following the same reasoning as in our analysis of the first component of the global variance in \cite{2025Hu}, this term coincides with that of \cite{2022Lin} under the balanced panel design.
However, in the second term, $\mathbb{D}_{ij}^{(2)}=\plim_{T\rightarrow \infty}T^{-1}\sum_{s=1}^T\sum_{t=1}^TE\left[\mathbf{G}_s\left(\lambda_{ij}^\prime\mathbf{F}_{is}+e_{ijs}\right)\left(\lambda_{ij}^\prime\mathbf{F}_{it}+e_{ijt}\right)\mathbf{G}_t^\prime\right]\geq\Omega_{\gamma, mi}$, therefore, \cite{2022Lin}'s estimation method yields a larger asymptotic variance for the estimated common components than ours.

\ \ \ \ As noted earlier, our penalized quasi-maximum-likelihood function gives a straightforward manner for deriving the central limit theorem for group factor models either in linear form or extended to nonlinear forms. 

\subsection{Computational issues}
We present two algorithms for numerically solving the objective function: an alternating least squares (ALS) algorithm, which is detailed in the Supplementary Material, and a gradient descent (GD) algorithm (Algorithm \ref{alg:GD}). Empirically in our simulations, we find the GD algorithm is computationally more efficient.
While ALS algorithm is more like those given in the literature for separately identifying the factors and their loadings, the GD algorithm is straightforwardly principled for our estimation approach via the Lagrangian trick. 

\begin{algorithm}[!ht]
    \caption{Gradient Descent (GD) for estimating group factor models}
    \label{alg:GD}
    \begin{algorithmic}[1]
        \STATE \textbf{Input:} Grouped data $\{ \mathbf{Y}_m \}_{m \leq M}$; the numbers of global and local factors $r_0$ and $r_m$ for $m=1,\dots,M$; learning rate $\eta>0$; maximum iterations $K_{\max}$.
        \STATE \textbf{Output:} Factor and loading matrices $\hat{\mathbf{\Gamma}}$, $\hat{\mathbf{\Lambda}}$, $\hat{\mathbf{G}}$ and $\hat{\mathbf{F}}$.
        \STATE let the initial estimators $\mathbf{\Gamma}^{(0)}$, $\mathbf{\Lambda}^{(0)}$, $\mathbf{G}^{(0)}$ and $\mathbf{F}^{(0)}$ by APM;
        \FOR{$k = 0,1,\ldots, K_{\max}-1$}
        \STATE compute the total loss $\mathbf{e}^{(k)} \leftarrow \sum_{m=1}^M \|\mathbf{Y}_m - \boldsymbol{\Gamma}_m^{(k)}\mathbf{G}^{(k)} - \boldsymbol{\Lambda}_m^{(k)}\mathbf{F}_m^{(k)}\|_F^2$.
        \FOR{$m=1$ to $M$}
            \STATE residual: $\mathbf{e}_m^{(k)} \leftarrow \mathbf{Y}_m - \boldsymbol{\Gamma}_m^{(k)}\mathbf{G}^{(k)} - \boldsymbol{\Lambda}_m^{(k)}\mathbf{F}_m^{(k)}$.
            \STATE gradients:
            $\nabla_{\boldsymbol{\Gamma}_m} \mathbf{L} \leftarrow - \mathbf{e}_m^{(k)} \mathbf{G}^{(k)\top}$,
            $\nabla_{\boldsymbol{\Lambda}_m} \mathbf{L} \leftarrow - \mathbf{e}_m^{(k)} \mathbf{F}_m^{(k)\top}$,
            $\nabla_{\mathbf{F}_m} \mathbf{L} \leftarrow -  \boldsymbol{\Lambda}_m^{(k)\top}\mathbf{e}_m^{(k)}$.
            \STATE update parameters:
            $\boldsymbol{\Gamma}_m^{(k+1)} \leftarrow \boldsymbol{\Gamma}_m^{(k)} - \eta \nabla_{\boldsymbol{\Gamma}_m}\mathbf{L}$,
            $\boldsymbol{\Lambda}_m^{(k+1)} \leftarrow \boldsymbol{\Lambda}_m^{(k)} - \eta \nabla_{\boldsymbol{\Lambda}_m}\mathbf{L}$,
            $\mathbf{F}_m^{(k+1)} \leftarrow \mathbf{F}_m^{(k)} - \eta \nabla_{\mathbf{F}_m}\mathbf{L}$.
        \ENDFOR
        \STATE accumulate gradient for global factors: $\nabla_{\mathbf{G}} \mathbf{L} \leftarrow -\sum_{m=1}^M  \boldsymbol{\Gamma}_m^{(k)\top}\mathbf{e}_m^{(k)}$.
        \STATE update global factors: $\mathbf{G}^{(k+1)} \leftarrow \mathbf{G}^{(k)} - \eta \nabla_{\mathbf{G}}\mathbf{L}$.
        \STATE compute average loss: $\bar{\mathbf{e}}^{(k+1)} \leftarrow \mathbf{e}^{(k+1)} / (T N)$.
        \STATE until convergence and output the estimators from the last step and denote them as $\hat{\mathbf{\Gamma}}$, $\hat{\mathbf{\Lambda}}$, $\hat{\mathbf{G}}$ and $\hat{\mathbf{F}}$.
        \ENDFOR
    \end{algorithmic}
\end{algorithm}

\section{Numerical Studies}\label{section4}
\subsection{Simulation Set Up}\label{Simulation Set Up}
Following the simulation design of \cite{2025Hu}, we generate the data from the following group factor model: 
\begin{equation*}
	y_{m,i,t} = \gamma'_{m,i} \mathbf{G}_t + \sqrt{h_{m1}} \lambda_{m,i}' \mathbf{F}_{m,t} + \sqrt{\kappa h_{m2}} e_{m,i,t},
\end{equation*}
for \(m=1,\ldots,M\), \(i=1,\ldots,N_m\), and \(t=1,\ldots,T\).

\ \ \ \ The global factors \(\mathbf{G}_t\), global loadings \(\gamma_{m,i}\), local loadings \(\lambda_{m,i}\), and idiosyncratic errors \(e_{m,i,t}\) are generated from the following data generating process: 
\begin{equation*}
	\mathbf{G}_t = \phi_G \mathbf{G}_{t-1} + \boldsymbol{w}_t^G, \quad \boldsymbol{w}_t^G \overset{\text{i.i.d.}}{\sim} N(0, \mathbf{I}_{r_0}),
\end{equation*}
\begin{equation*}
	\gamma_{m,i} \overset{\text{i.i.d.}}{\sim} N(0,\mathbf{I}_{r_0}), \quad \lambda_{m,i} \overset{\text{i.i.d.}}{\sim} N(0,\mathbf{I}_{r_m}),
\end{equation*}
\begin{equation*}
	e_{m,i,t} = \phi_e e_{m,i,t-1} + v_{m,i,t} + \beta \sum_{1 \leq |h| \leq 8} v_{m,i-h,t}, \quad v_{m,i,t} \overset{\text{i.i.d.}}{\sim} N(0, 1).
\end{equation*}
where the global and local factors are temporally autocorrelated, and the idiosyncratic errors are allowed to be both serially and cross-sectionally correlated.

\ \ \ \ To thoroughly evaluate the robustness of the proposed method against correlated local structures, we consider the following three scenarios for the local factors \(\mathbf{F}_{m,t}\):

\textbf{Case 1 (Independent local factors):}
\begin{equation*}
	\mathbf{F}_{m,t} = \phi_F \mathbf{F}_{m,t-1} + \mathbf{w}_{m,t}^F, \quad \mathbf{w}_{m,t}^F \overset{\text{i.i.d.}}{\sim} N(\mathbf{0}, \mathbf{I}_{r_m}).
\end{equation*}

\textbf{Case 2 (Common local factors across subgroups):}

\ \ \ \ Assume $M$ is even, and define two base processes:
\begin{align*}
	\mathbf{F}_t^1 &= \phi_F \mathbf{F}_{t-1}^1 + \mathbf{w}_t^{F1}, \quad \mathbf{w}_t^{F1} \overset{\text{i.i.d.}}{\sim} N(\mathbf{0}, \mathbf{I}_{\tilde{r}}), \\
	\mathbf{F}_t^2 &= \phi_F \mathbf{F}_{t-1}^2 + \mathbf{w}_t^{F2}, \quad \mathbf{w}_t^{F2} \overset{\text{i.i.d.}}{\sim} N(\mathbf{0}, \mathbf{I}_{\tilde{r}}),
\end{align*}
and set $\mathbf{F}_{1,t} = \dots = \mathbf{F}_{M/2, t} = \mathbf{F}_t^1$ and $\mathbf{F}_{M/2 + 1, t} = \dots = \mathbf{F}_{M,t} = \mathbf{F}_t^2$, where $r_1 = \dots = r_M := \tilde{r}$.

\textbf{Case 3 (Cross-group correlated local factors):}
\begin{equation*}
	\mathbf{F}_t = \phi_F \mathbf{F}_{t-1} + \mathbf{w}_t^F, \quad \mathbf{w}_t^F \overset{\text{i.i.d.}}{\sim} N(\mathbf{0}, \boldsymbol{\Sigma}_F),
\end{equation*}
where $\boldsymbol{\Sigma}_F$ is an $\sum_{m=1}^M r_m \times \sum_{m=1}^M r_m$ matrix with unit diagonal elements, and off-diagonal elements all equal to $0.5$.

\ \ \ \ To balance the variance contributions of the global, local, and idiosyncratic components when \(\kappa=1\), we set
\begin{equation*}
	h_{m1} = \left( \frac{r_0}{1 - \phi_G^2} \right) \Big/ \left( \frac{r_m}{1 - \phi_F^2} \right), \quad
	h_{m2} = \left( \frac{r_0}{1 - \phi_G^2} \right) \Big/ \left( \frac{1 + 16\beta^2}{1 - \phi_e^2} \right).
\end{equation*}
where \(r_0 = 0\), we set \(h_{m1} = 1\) and correct the variance of the idiosyncratic errors by setting $h_{m2} = \left( \frac{r_m}{1 - \phi_G^2} \right) \Big/ \left( \frac{1 + 16\beta^2}{1 - \phi_e^2} \right)$ to maintain equal contribution across the remaining components.

\ \ \ \ We consider a comprehensive set of sample size configurations. For \(T=50\), we take either balanced group sizes \(N_m \in \{30, 50\}\) for all \(m\), or unbalanced sizes with \(N_m \overset{\text{i.i.d.}}{\sim} \text{DUnif}[30, 50]\) (denoted as \(U_1\)), where DUnif denotes the Discrete Uniform Distribution.
Similarly, for \(T=100\), we use \(N_m \in \{50, 100\}\) or \(N_m \overset{\text{i.i.d.}}{\sim} \text{DUnif}[50,100]\) (denoted as \(U_2\)); for \(T=200\), we use \(N_m \in \{100, 200\}\) or \(N_m \overset{\text{i.i.d.}}{\sim} \text{DUnif}[100,200]\) (denoted as \(U_3\)). The number of groups is set to \(M \in \{4, 10\}\), the noise level to \(\kappa \in \{1, 3\}\), and the numbers of global factors to \(r_0 \in \{0, 2\}\) and \(r_1 = \cdots = r_M = 2\). The remaining hyperparameters are fixed at \(\phi_G = \phi_F = \phi_e = 0.5\) and \(\beta = 0.1\). The maximum number of group factors considered in the estimation is set to \(r_{\max} = r_0 + r_m + 4\). For each configuration, we conduct \(500\) independent replications and assess the estimation accuracy via the reconstruction error of the original data using the estimated factor spaces.

\subsection{Comparison}\label{sec4.2}
Table \ref{table1} summarizes the average Mean Squared Error (MSE) between the true data $\mathbf{Y}$ and the reconstructed data $\hat{\mathbf{Y}}$ based on the estimated group factor model, with $500$ replications for the configuration $r_0=2, M=4$. The estimates of loadings and factors are obtained via our proposed Gradient Descent (GD) algorithm. To benchmark its performance, we compare against several representative existing approaches, including the Canonical Correlation Difference (CCD) and Modified Canonical Correlation (MCC) of \cite{2023Choi}, the Circular Projection Estimation (CPE) of \cite{2023Chen}, the Generalised Canonical Correlation (GCC) of \cite{2022Lin}, and the Aggregated Projection Method (APM) of \cite{2025Hu}. Additional simulation result for the case $r_0=2, M=10$ is reported in the supplementary material. Our GD-based method exhibits superior and robust performance over all competitors, irrespective of the noise level or the correlation structure among local factors.

\begin{table}
	\centering
	\caption{The accuracy of the estimates when $r_0=2, M=4$.}
    \label{table1}
	\footnotesize
    \setlength{\tabcolsep}{4pt}
    \renewcommand{\arraystretch}{0.8}
	\begin{tabular}[t]{c c c c c c c c c c c c c c}
		\hline
		\multicolumn{2}{c}{} & \multicolumn{6}{c}{$\kappa=1$} & \multicolumn{6}{c}{$\kappa=3$} \\
\cmidrule(lr){3-8} \cmidrule(lr){9-14}
T & $N$ & CCD & MCC & CPE & GCC & APM & GD & CCD & MCC & CPE & GCC & APM & GD \\
        \hline
        \multicolumn{2}{c}{case 1 } \\
        50 & 30 & 1.454 & 1.454 & 1.470 & 1.450 & 1.450 & \textbf{1.438} & 2.291 & 2.291 & 2.318 & 2.279 & 2.278 & \textbf{2.256}\\
        50 & $U_1$ & 1.491 & 1.491 & 1.508 & 1.490 & 1.489 & \textbf{1.481} & 2.376 & 2.376 & 2.403 & 2.367 & 2.365 & \textbf{2.345}\\
        50 & 50 & 1.509 & 1.509 & 1.519 & 1.507 & 1.506 & \textbf{1.500} & 2.476 & 2.476 & 2.494 & 2.468 & 2.467 & \textbf{2.451}\\
        100 & 50 & 1.550 & 1.550 & 1.562 & 1.549 & 1.549 & \textbf{1.543} & 2.566 & 2.566 & 2.586 & 2.562 & 2.561 & \textbf{2.549}\\
        100 & $U_2$ & 1.565 & 1.565 & 1.577 & 1.564 & 1.564 & \textbf{1.561} & 2.645 & 2.645 & 2.660 & 2.640 & 2.640 & \textbf{2.632}\\
        100 & 100 & 1.576 & 1.576 & 1.585 & 1.575 & 1.575 & \textbf{1.573} & 2.705 & 2.705 & 2.726 & 2.703 & 2.703 & \textbf{2.698}\\
        200 & 100 & 1.594 & 1.594 & 1.603 & 1.593 & 1.593 & \textbf{1.592} & 2.745 & 2.745 & 2.765 & 2.744 & 2.744 & \textbf{2.740}\\
        200 & $U_3$ & 1.602 & 1.602 & 1.607 & 1.602 & 1.602 & \textbf{1.601} & 2.766 & 2.766 & 2.780 & 2.766 & 2.766 & \textbf{2.764}\\
        200 & 200 & \textbf{1.604} & \textbf{1.604} & 1.610 & 1.605 & 1.605 & \textbf{1.604} & \textbf{2.773} & \textbf{2.773} & 2.786 & 2.774 & 2.774 & \textbf{2.773}\\ 
        \multicolumn{2}{c}{case 2 } \\
        50 & 30 & 1.529 & 1.529 & 1.469 & 1.456 & 1.449 & \textbf{1.436} & 2.310 & 2.310 & 2.312 & 2.274 & 2.274 & \textbf{2.246}\\
        50 & $U_1$ & 1.566 & 1.566 & 1.499 & 1.490 & 1.481 & \textbf{1.471} & 2.391 & 2.391 & 2.387 & 2.353 & 2.354 & \textbf{2.328}\\
        50 & 50 & 1.598 & 1.598 & 1.519 & 1.516 & 1.506 & \textbf{1.499} & 2.505 & 2.505 & 2.492 & 2.467 & 2.466 & \textbf{2.446}\\
        100 & 50 & 1.654 & 1.654 & 1.559 & 1.558 & 1.546 & \textbf{1.541} & 2.613 & 2.613 & 2.586 & 2.563 & 2.562 & \textbf{2.547}\\
        100 & $U_2$ & 1.669 & 1.669 & 1.570 & 1.574 & 1.562 & \textbf{1.559} & 2.720 & 2.720 & 2.687 & 2.669 & 2.662 & \textbf{2.655}\\
        100 & 100 & 1.688 & 1.688 & 1.583 & 1.587 & 1.574 & \textbf{1.572} & 2.761 & 2.761 & 2.723 & 2.707 & 2.700 & \textbf{2.696}\\
        200 & 100 & 1.726 & 1.726 & 1.603 & 1.609 & 1.594 & \textbf{1.592} & 2.809 & 2.809 & 2.764 & 2.750 & 2.742 & \textbf{2.738}\\
        200 & $U_3$ & 1.719 & 1.719 & 1.610 & 1.617 & 1.603 & \textbf{1.602} & 2.838 & 2.838 & 2.783 & 2.776 & 2.768 & \textbf{2.766}\\
        200 & 200 & 1.706 & 1.706 & 1.610 & 1.619 & \textbf{1.604} & \textbf{1.604} & 2.834 & 2.834 & 2.786 & 2.782 & 2.774 & \textbf{2.773}\\
        \multicolumn{2}{c}{case 3 } \\
        50 & 30 & 1.394 & 1.394 & 1.410 & 1.390 & 1.389 & \textbf{1.379} & 2.201 & 2.201 & 2.232 & 2.193 & 2.192 & \textbf{2.170}\\
        50 & $U_1$ & 1.420 & 1.420 & 1.437 & 1.417 & 1.416 & \textbf{1.407} & 2.343 & 2.343 & 2.379 & 2.339 & 2.337 & \textbf{2.319}\\
        50 & 50 & 1.476 & 1.476 & 1.488 & 1.475 & 1.474 & \textbf{1.468} & 2.397 & 2.397 & 2.417 & 2.391 & 2.390 & \textbf{2.375}\\
        100 & 50 & 1.521 & 1.521 & 1.533 & 1.520 & 1.520 & \textbf{1.515} & 2.474 & 2.474 & 2.496 & 2.470 & 2.470 & \textbf{2.458}\\
        100 & $U_2$ & 1.559 & 1.559 & 1.572 & 1.558 & 1.558 & \textbf{1.555} & 2.602 & 2.602 & 2.621 & 2.599 & 2.599 & \textbf{2.593}\\
        100 & 100 & 1.570 & 1.570 & 1.579 & 1.570 & 1.570 & \textbf{1.568} & 2.647 & 2.647 & 2.666 & 2.645 & 2.645 & \textbf{2.641}\\
        200 & 100 & 1.591 & 1.591 & 1.600 & 1.590 & 1.590 & \textbf{1.589} & 2.688 & 2.688 & 2.708 & 2.687 & 2.687 & \textbf{2.684}\\
        200 & $U_3$ & \textbf{1.600} & \textbf{1.600} & 1.606 & \textbf{1.600} & \textbf{1.600} & \textbf{1.600} & \textbf{2.752} & \textbf{2.752} & 2.764 & 2.753 & 2.753 & \textbf{2.752}\\
        200 & 200 & \textbf{1.603} & \textbf{1.603} & 1.609 & 1.604 & 1.604 & \textbf{1.603} & \textbf{2.764} & \textbf{2.764} & 2.776 & 2.765 & 2.765 & \textbf{2.764}\\
        \hline
    \end{tabular}
\end{table}

\subsection{Asymptotic distribution}
\begin{figure}[ht]
	\centering
	\includegraphics[width=1\textwidth]{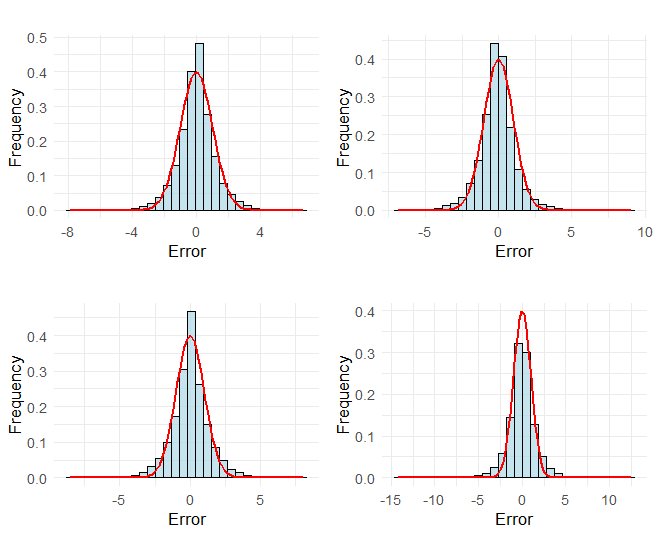}
	\caption{Histograms of the standardized estimators: $\hat{\gamma}_{m,i}$ (top-left), $\hat{\lambda}_{m,i}$ (top-right), $\hat{\mathbf{G}}_{t}$ (bottom-left) and $\hat{\mathbf{F}}_{m,t}$ (bottom-right) over 100 replications. The red solid curve represents the standard normal distribution.}
    \label{fig1}
\end{figure}

In this section, we assess the accuracy of the asymptotic distribution derived in Theorem \ref{theorem2}. Specifically, we adopt Case 1 with $T=100$, $N_1=\dots=N_M=50$, $M=4$, $r_0=2$, and $\kappa=1$. The data generation and standardization follow the procedure detailed in Section \ref{Simulation Set Up}.

\ \ \ \ Theorem \ref{theorem2} establishes the asymptotic normality of the estimators: $\hat{\gamma}_{m,i} - \gamma_{m,i}^0 \rightarrow N\left(\mathbf{0},\hat{\text{var}}_{\gamma}\right)$, $\hat{\lambda}_{m,i} - \lambda_{m,i}^0 \rightarrow N\left(\mathbf{0},\hat{\text{var}}_{\lambda}\right)$, $\hat{\mathbf{G}}_{t} - \mathbf{G}_{t}^0 \rightarrow N\left(\mathbf{0},\hat{\text{var}}_g\right)$ and $\hat{\mathbf{F}}_{m,t} - \mathbf{F}_{m,t}^0 \rightarrow N\left(\mathbf{0},\hat{\text{var}}_f\right)$. Figure \ref{fig1} shows the normalized histograms of $\hat{\text{var}}_{\gamma}^{-1/2}\left(\hat{\gamma}_{m,i} - \gamma_{m,i}^0 \right)$, $\hat{\text{var}}_{\lambda}^{-1/2}\left(\hat{\lambda}_{m,i} - \lambda_{m,i}^0\right)$, $\hat{\text{var}}_g^{-1/2}\left(\hat{\mathbf{G}}_{t} - \mathbf{G}_{t}^0\right)$ and $\hat{\text{var}}_f^{-1/2}\left(\hat{\mathbf{F}}_{m,t} - \mathbf{F}_{m,t}^0\right)$, overlaid with the standard normal density curve.
In each panel, the empirical distribution closely matches the theoretical normal density, confirming the validity of the asymptotic approximation.

\section{Empirical Applications}\label{section5}
\subsection{US housing price growth rate}
In this section, we revisit the U.S. housing price growth rate dataset analyzed in the \cite{2025Hu}, but employ a newly proposed one-step estimation procedure for the group factor model. 
In contrast to the two-step APM, which first performs factor analysis separately within each group and then disentangles the global and local factors via aggregated projection and subsequent PC estimation, the one-step approach directly estimates all factors and loadings jointly by optimizing a unified objective function. 

\ \ \ \ We apply the one-step method to the same dataset, which consists of monthly Zillow Home Value Index (ZHVI) growth rates for 16 U.S. states from January 2000 to April 2023. The data comprise \(M = 16\) groups, \(N = \sum_{m=1}^M N_m = 2241\) regions, and \(T = 279\) time points. The number of global factors is determined by the eigenvalue gap criterion, which selects one global factor. The estimated global factor explains a substantial proportion of the total variation, with an average relative importance of \(56.1\%\) across states.

\begin{figure}[ht]
	\centering
	\includegraphics[width=1\textwidth]{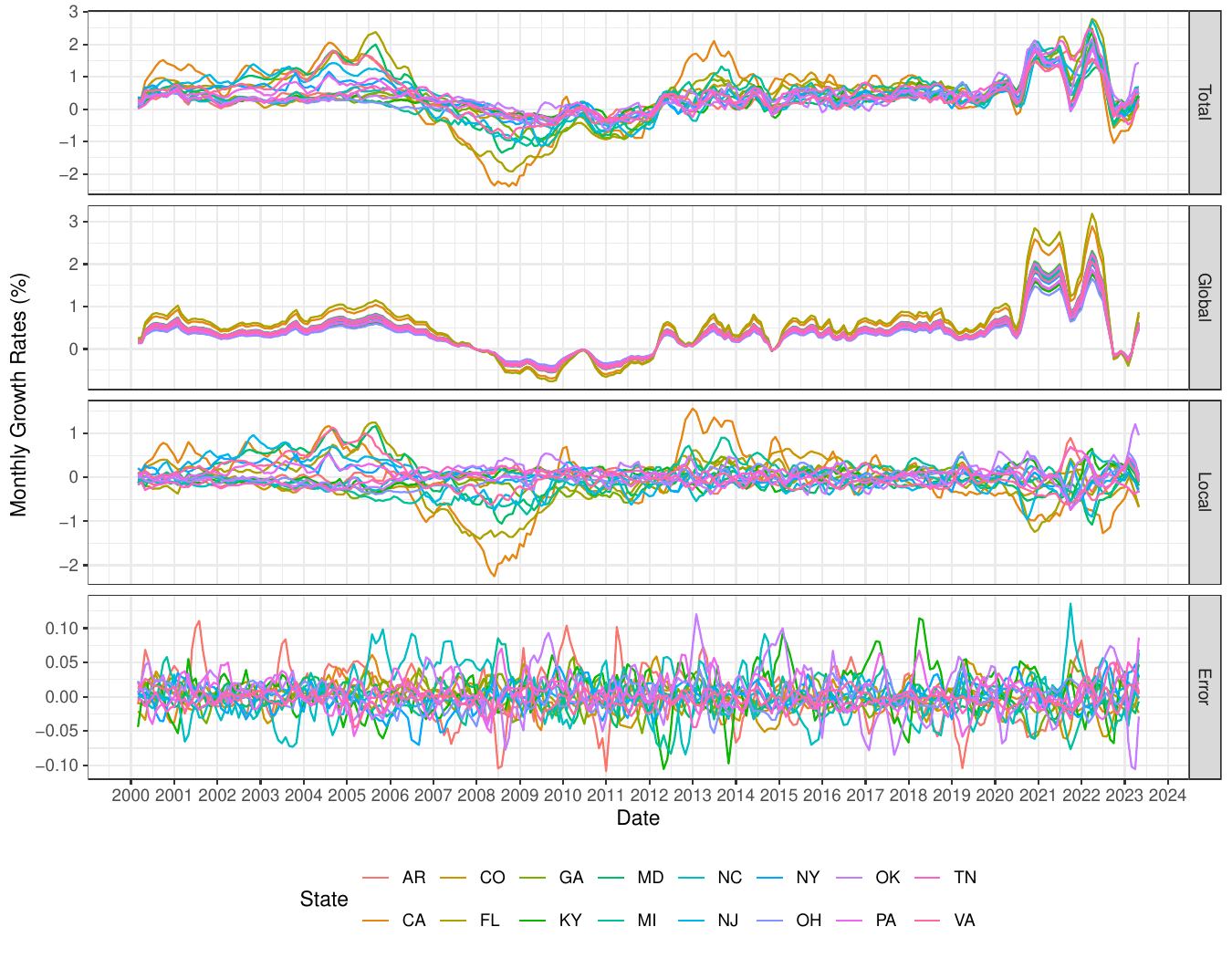}
	\caption{The estimated global, local and error components in each state.}
    \label{fig2}
\end{figure}

\begin{table}[!htb]
\centering
\caption{Table of the summarized statistics and the estimation results of GD algorithm}
\label{GDstatistics}
\begin{tabular}{@{\extracolsep{5pt}} cccccccc}
\hline
State & $N_m$ & mean & sd & $\hat{r}_m$ & $\text{RIG}_m$ & $\text{RIF}_m$ & $\text{RIE}_m$ \\
\hline
AR & 90 & 0.353 & 0.607 & 2 & 0.578 & 0.106 & 0.316 \\
CA & 114 & 0.457 & 1.176 & 5 & 0.441 & 0.468 & 0.092 \\
CO & 107 & 0.427 & 0.688 & 5 & 0.588 & 0.266 & 0.145 \\
FL & 206 & 0.460 & 1.078 & 5 & 0.621 & 0.280 & 0.099 \\
GA & 232 & 0.350 & 0.795 & 4 & 0.610 & 0.170 & 0.220 \\
KY & 93 & 0.332 & 0.629 & 2 & 0.526 & 0.135 & 0.340 \\
MD & 95 & 0.361 & 0.804 & 5 & 0.555 & 0.327 & 0.118 \\
MI & 123 & 0.328 & 0.841 & 4 & 0.479 & 0.267 & 0.253 \\
NC & 99 & 0.368 & 0.728 & 3 & 0.616 & 0.142 & 0.241 \\
NJ & 91 & 0.384 & 0.778 & 5 & 0.581 & 0.314 & 0.105 \\
NY & 250 & 0.404 & 0.658 & 5 & 0.580 & 0.206 & 0.214 \\
OH & 213 & 0.276 & 0.598 & 4 & 0.531 & 0.186 & 0.283 \\
OK & 94 & 0.398 & 0.755 & 3 & 0.418 & 0.231 & 0.351 \\
PA & 105 & 0.365 & 0.578 & 4 & 0.619 & 0.166 & 0.215 \\
TN & 205 & 0.420 & 0.661 & 2 & 0.698 & 0.084 & 0.218 \\
VA & 124 & 0.386 & 0.720 & 6 & 0.535 & 0.334 & 0.130 \\
\hline
\end{tabular}
\end{table}


\ \ \ \ Table \ref{GDstatistics} reports the summary statistics and the estimation results obtained using our GD algorithm. In the table, $N_m$, mean, and sd denote the number of regions in each state, the mean and the standard deviation of $y_{m,i,t}$. The column $\hat{r}_m$ gives the estimated number of local factors for each state. Additionally, for each state $m$, we compute the variance proportions attributed to the global factors, local factors, and idiosyncratic errors following the definitions in \citet{2025Hu}:
\begin{align*} 
\text{RIG}_m&=\frac{1}{N_m}\sum_{i=1}^{N_m}\frac{\widehat{\mathbf{\gamma}}_{m,i}'\widehat{\mathbf{\gamma}}_{m,i}}{T^{-1}\mathbf{y}_{m,i}'\mathbf{y}_{m,i}},\\ 
\text{RIF}_m&=\frac{1}{N_m}\sum_{i = 1}^{N_m}\frac{\widehat{\mathbf{\lambda}}_{m,i}'\widehat{\mathbf{\lambda}}_{m,i}}{T^{-1}\mathbf{y}_{m,i}'\mathbf{y}_{m,i}}, \\
\text{RIE}_m& = 1-\text{RIG}_m-\text{RIF}_m .
\end{align*}

\ \ \ \ On average, the global factor explains approximately $56\%$ of the total variation across states under the GD estimation, with $\text{RIG}_m$ ranging from $41.8\%$ in Oklahoma (OK) to $69.8\%$ in Tennessee (TN). The local factors account for a moderate proportion of the variation, while the remaining variation is attributed to idiosyncratic errors. Corresponding results obtained using the APM method are reported in Table A.2 of the supplementary material.

\ \ \ \ The one-step estimation yields a clean decomposition of the housing price growth into three interpretable components at the state level, as visualized in Figure \ref{fig2}. The global component faithfully mirrors the well-known national housing market cycles, including the downturn during the 2008 financial crisis and the sharp rebound amid the 2020–2021 pandemic. In contrast, the local components capture the remaining state-specific patterns, highlighting regional heterogeneities that are not attributable to the common national factor.

\ \ \ \ To further evaluate the performance of the GD algorithm relative to other existing methods, we apply the GD, CCD, MCC, CPE, GCC, and APM algorithms described in Section \ref{sec4.2} to estimate the latent factors driving the U.S. state-level housing price growth rates, as well as the corresponding residuals. Let $\mathbf{Y}$ denote the observed log-growth rate matrix of housing prices, and $\hat{\mathbf{Y}}$ denote the reconstructed matrix obtained from each method. We define the mean squared error (MSE) and the unexplained proportion of total variance as
\begin{equation*}
    \text{MSE}=\frac{1}{2241\times279}\|\hat{\mathbf{Y}}-\mathbf{Y}\|_F^2,
    \quad
    \rho=\frac{\|\hat{\mathbf{Y}}-\mathbf{Y}\|_F^2}{\|\mathbf{Y}-\Bar{\mathbf{Y}}\|_F^2},
\end{equation*}
respectively, where $\bar{\mathbf{Y}}$ denotes the column-wise mean matrix of $\mathbf{Y}$.

\ \ \ \ Table \ref{table5} reports the MSE and $\rho$ values for all competing methods, including CCD, MCC, CPE, GCC, APM, and our proposed one-step approach (GD). In addition, \(r_0\) denotes the number of global factors estimated by each method, and \(\bar{r}\) is the estimated average number of local factors. The proposed GD approach achieves the lowest MSE and $\rho$, demonstrating its clear superiority over the existing alternatives.

\begin{table}[!h]
\centering
\caption{Comparison of MSE and proportion of unexplained variance among six methods}
\label{table5}
\centering
\begin{tabular}[t]{ccccc}
\hline
Method & MSE & $\rho$ & $r_0$ & \(\bar{r}\)\\
\hline
GD & $\mathbf{0.1385}$ & $\mathbf{0.2326}$ & 1 & 4.00\\
APM & 0.1388 & 0.2331 & 1 & 4.00\\
GCC & 0.1390 & 0.2336 & 1 & 4.00\\
CPE & 0.1393 & 0.2340 & 1 & 4.00\\
CCD & 0.1423 & 0.2392 & 1 & 3.69\\
MCC & 0.1423 & 0.2392 & 1 & 3.69\\
\hline
\end{tabular}
\end{table}

\subsection{Weekly returns of CSI300 dataset}
In the second real data example, we employ the proposed method to analyze the returns of constituents of the China CSI 300 Index. We collect the daily closing price series for 300 stocks over a five-year period from 2021 to 2025. These stocks are traded on the Shanghai Stock Exchange and the Shenzhen Stock Exchange, respectively. To reduce serial dependence, we compute the weekly returns for each stock based on the closing price of the last trading day of each week, yielding a total of 261 weeks. Given that some stocks have relatively short listing histories and contain numerous missing observations, we retain only 274 stocks of which the missing rate is less than 5\%, and impute the missing values using linear interpolation (the overall missing rate is 1.97\%). The imputed weekly return series are further standardized to have zero mean and unit variance.  Finally, a panel data set of dimensions $261 \times 274$ is constructed, in which the 274 stocks are naturally split into two groups according to their listing exchanges:  169 stocks listed on the Shanghai (SH) Stock Exchange and 105 stocks listed on the Shenzhen (SZ) Stock Exchange.

\ \ \ \ The first step is to determine the numbers of global and local factors. The information criterion proposed in \cite{2025Hu} reports 4 global factors, 4 local factors for the Shanghai Stock Exchange and 1 local factor for the Shenzhen  Stock Exchange. Table \ref{tab:csi300} reports the proportions of explained variance by the global ($\text{RIG}_m$) and local ($\text{RIF}_m$) factors, and of the unexplained variance in the errors  ($\text{RIE}_m$), respectively for the two stock exchanges and by two methods, GD and APM.  The global factors from the proposed GD method explain slightly more variance of the returns on Shenzhen Stock Exchange than those from the APM method, but explain less variability for the Shanghai Stock Exchange. By contrast, the local factors from the proposed GD method explain slightly more variance of the returns on Shanghai Stock Exchange, and show comparable performance on the Shenzhen Stock Exchange.  Overall, the two methods lead to very close estimation.

\begin{figure}[ht]
	\centering
	\includegraphics[width=1\textwidth]{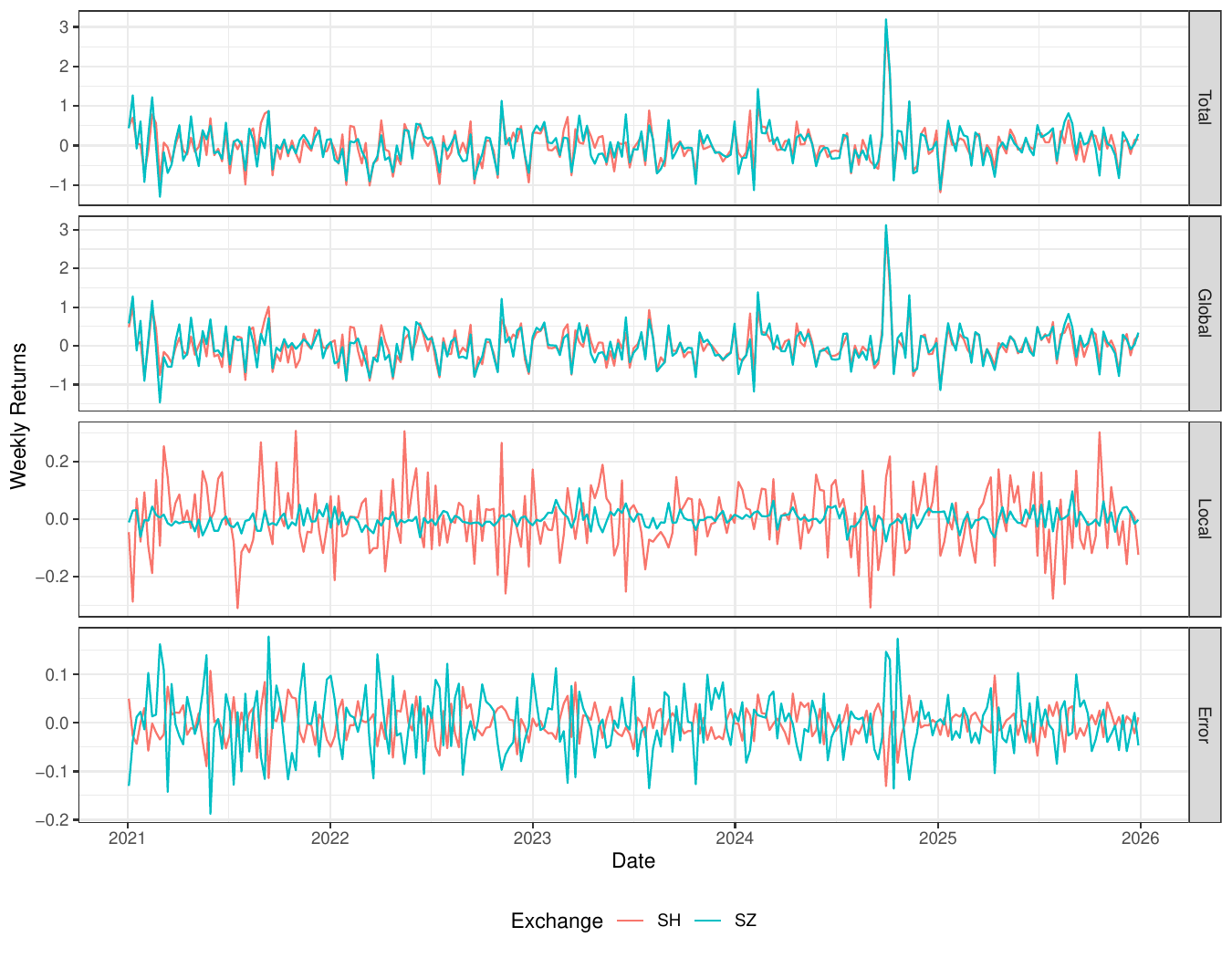}
	\caption{The estimated global, local and error components for the two stock exchanges. }
    \label{fig: plot csi300}
\end{figure}

\begin{table}[!ht]
\centering
\caption{Table of the summarized statistics for CSI300 data set}
\label{tab:csi300}
\begin{tabular}{@{\extracolsep{5pt}} cccccccccc}
\hline
\multirow{2}{*}{Exchange} & \multirow{2}{*}{$N_m$} & \multirow{2}{*}{$\hat{r}_m$} & \multicolumn{3}{c}{GD} & \multicolumn{3}{c}{APM} \\
\cmidrule(lr){4-6} \cmidrule(lr){7-9}
 & & & $\text{RIG}_m$ & $\text{RIF}_m$ & $\text{RIE}_m$ & $\text{RIG}_m$ & $\text{RIF}_m$ & $\text{RIE}_m$ \\
\hline
SH & 169 & 4 & 0.3682 & 0.1235 & 0.5082 & 0.3721 & 0.1228 & 0.5051 \\
SZ & 105 & 1 & 0.3820 & 0.0369 & 0.5811 & 0.3786 & 0.0369 & 0.5844 \\
\hline
\end{tabular}
\end{table}

\ \ \ \ Moreover, Figure \ref{fig: plot csi300} plots the estimated series of the total common components, the common components related to the global factors,  the common components related to the local factors, and the idiosyncratic error series, respectively on the two exchanges and averaged over all the stocks, by the GD method. It's seen that the local factors perform quite differently for the two stock exchanges: those from the Shanghai exchange show more variability, while those from the Shenzhen exchange are more stable. In other words, the proposed method successfully extracts the local features of the stock market.

\ \ \ \ Finally, we further  develop a rolling validation procedure to compare different estimators. For each week $t$, we repeatedly fit a group factor model based on the $n$-week's (bandwidth) observations before (including) $t$ by different methods, and record the mean squared error (in-sample MSE). Then, we take the estimated global and local loadings as explanatory variables, and take the observations in $(t+1)$-th week as the response variable. We fit a linear regression model and report the associated mean squared error (out-of-sample MSE). Table \ref{table: mse csi300} shows the in-sample and out-of-sample MSEs for four methods with different bandwidth $n$, where the numbers of group and local factors are set to be the same, i.e., $r_0=4, r_1=4,r_2=1$. It's seen that the proposed GD method always leads to the lowest in-sample MSE, and also leads to the lowest or comparable out-of-sample MSE in most cases.

\begin{table}[htb]
\centering
\caption{Comparison of in-sample and out-of-sample MSEs in the rolling validation process among 4 methods for CSI300 data set.}
\label{table: mse csi300}
\centering
\begin{tabular}[t]{lccccccccccc}
\hline
 & & \multicolumn{4}{c}{In-sample MSE} & & \multicolumn{4}{c}{Out-of-sample MSE} \\
\cmidrule(lr){3-6} \cmidrule(lr){8-11}
$n$ & & APM & GD & CPE & GCC & & APM & GD & CPE & GCC \\
\hline
50&&0.4050& $\mathbf{0.4017}$&0.4076&0.4029&&0.5015&0.5012&0.5018&$\mathbf{0.5010}$\\
75&&0.4307& $\mathbf{0.4290}$&0.4355&0.4310&&0.4815&$\mathbf{0.4813}$&0.4823&$\mathbf{0.4813}$\\
100&&0.4431& $\mathbf{0.4416}$&0.4480&0.4437&&0.4613&$\mathbf{0.4610}$&0.4618&0.4615\\
125&&0.4561& $\mathbf{0.4546}$&0.4611&0.4567&&0.4434&0.4434&$\mathbf{0.4433}$&0.4437\\
150&&0.4652& $\mathbf{0.4638}$&0.4701&0.4659&&0.4424&$\mathbf{0.4423}$&0.4437&0.4427\\
\hline
\end{tabular}
\end{table}

\section{Conclusion}\label{section6}
This paper proposes a one-step penalized quasi-likelihood estimation method for group factor models. Under mild regularity conditions, we establish the consistency, convergence rates, and asymptotic normality of the estimated factors and loadings. A carefully designed penalized loss function enables simultaneous and consistent estimation of all model parameters. Extensive simulation studies demonstrate that the proposed method outperforms existing approaches. The method is further applied to two empirical settings: modeling the growth rate of housing prices across U.S. states and analyzing the returns of constituents of the China CSI 300 Index.

\section*{Supplementary Materials}
The supplementary materials contain the details of the proof, an additional algorithm, and further simulation and empirical results.

\section*{Disclosure Statement} The authors report there are no competing interests to declare.

\section*{Acknowledgement} Kong's work is partially supported by NSFC (72342019, 12431009). Pan's research is partially supported by Postgraduate Research \& Practice Innovation Program of Jiangsu Province (KYCX25\_2455). The first and corresponding authors are partially supported by the Priority Academic Program Development of Jiangsu Higher Education Institutions (Statistics) and the Open Project of Joint Lab for Statistics and Finance of NAU (2025JLSF101).

\bibliographystyle{chicago}
\bibliography{pxyref}

\end{document}